\documentclass[a4paper,11pt]{article}

\usepackage{amsfonts, amsmath, amssymb}
\usepackage[utf8]{inputenc}
\usepackage[a4paper,top=3cm,bottom=2.5cm,left=2.5cm,right=2.5cm, bindingoffset=5mm]{geometry}
\usepackage{color}
\usepackage{booktabs}
\usepackage{subfigure}
\usepackage{graphicx}% include figures
\usepackage{float}
\usepackage{hyperref}
\hypersetup{citecolor=blue}
\usepackage{array}
\usepackage{colordvi}
\begin{document}

\title{\bf Warm-assisted natural inflation}
\author{Yakefu Reyimuaji$^{a}$\footnote{yakefu@mail.itp.ac.cn}, Xinyi Zhang$^{b,c}$\footnote{corresponding author: zhangxinyi@ihep.ac.cn}, }
\date{
$^a${\em \small CAS Key Laboratory of Theoretical Physics, Institute of Theoretical Physics, \\
Chinese Academy of Sciences, Beijing 100190, China} \\
$^b${\em \small Institute of High Energy Physics, Chinese Academy of Sciences, Beijing 100049, China}\\
$^c${\em \small School of Physics and State Key Laboratory of Nuclear Physics and Technology, \\ Peking University, Beijing 100871, China}
 }
\maketitle

\begin{abstract}
We consider natural inflation in a warm inflation framework with a temperature-dependent dissipative coefficient $\Gamma \propto  T^3$. Natural inflation can be compatible with the Planck 2018 results with such warm assistance. With no a priori assumptions on the dissipative effect's magnitude, we find that the Planck results prefer a weak dissipative regime for our benchmark scale $f=5 M_{\rm pl}$, which lies outside the $2\sigma$ region in the cold case. The inflation starts in the cold regime and evolves with a growing thermal fluctuation that dominates over quantum fluctuation before the end of the inflation. The observed spectral tilt puts stringent constraints on the model's parameter space. We find that $f< 1 M_{\rm pl}$ is excluded. A possible origin of such dissipative coefficient from axion-like coupling to gauge fields and tests of the model are also discussed.
\end{abstract}

\section{Introduction}
The big bang theory offers successful explanations to various observations but requires delicate initial conditions. Inflation explains the problems regarding initial conditions by offering an early-stage accelerating phase~\cite{Guth:1980zm}. To have enough inflation that matches the observation, the inflation potential has to be very flat. Such flatness usually requires some fine-tuning unless a symmetry protect it. Natural inflation~\cite{Freese:1990rb,Adams:1992bn} provides a good explanation of the flatness of the inflation potential. In natural inflation models, the inflaton is an axion-like particle whose potential is protected by a shift symmetry. 

Though well-motivated, natural inflation is strongly disfavored by Planck 2018 data~\cite{Akrami:2018odb}. Besides, the observation favors a large $f$ ($f<5.4 M_{\rm pl}$ is disfavored by Planck 2018 data at the $2\sigma$ level. Here $M_{\rm pl}$ is the reduced Planck mass, and $f$ is the scale at which shift symmetry is spontaneously breaking). It is known that $f>1 M_{\rm pl}$ may lead to large corrections in action; thus it is difficult to be realized in a more fundamental theory~\cite{Martin:2013tda}.

Inflaton has to couple to other particles to source reheating. In cold inflation models, this coupling is not taken into consideration until the end of inflation. While warm inflation~\cite{Berera:1995wh,Berera:1995ie} considers the role of this coupling during the inflation period and particles are produced concurrently. When the relativistic particles thermalize fast enough, they can be viewed as radiation. Inflaton's interaction with the radiation bath causes further energy loss as the radiation bath now contributes thermal friction besides the Hubble friction. The thermal friction modifies both the background evolution and the inflaton evolution. We consider natural inflation in a warm inflation scenario, which we call warm natural inflation, to see if:
\begin{enumerate}
  \item warm natural inflation can be compatible with the observations;
  \item the allowed parameter space includes $f<1 M_{\rm pl}$. 
\end{enumerate}

Natural inflation has been studied in warm inflation scenarios in Refs.~\cite{Mohanty:2008ab,Visinelli:2011jy,Mishra:2011vh}, where the dissipative coefficient is taken to be independent of temperature, and the calculation is performed in a strong dissipative regime. We consider here that the dissipative coefficient has a temperature dependence, which is also the case investigated in Refs.~\cite{Visinelli:2014qla,Visinelli:2016rhn,Wang:2019ozs,Ghadiri:2018nok,Berghaus:2019whh} for other inflationary models. To be more specific, we assume a dissipative coefficient with temperature dependence as $\Gamma \propto T^3$, which is justified from an axion-like interaction with gauge fields in strong dissipative regime~\cite{Berghaus:2019whh}. We proceed without further assumption on the strength of the dissipative effect. We find that natural inflation can be brought into the Planck-allowed range with a weak dissipative effect. Our parameter space is very constrained, which signals high testability of the model. More importantly, $f<1 M_{\rm pl}$ is excluded which differs from results in Refs.~\cite{Mohanty:2008ab,Visinelli:2011jy,Mishra:2011vh}.

We organize the paper as follows. In Section~\ref{sec:model} we introduce the generic features of warm inflation and then introduce warm natural inflation. We present our main results in Section~\ref{sec:result} and conclude in Section~\ref{sec:conclusion}.

\section{Warm inflation and warm natural inflation}\label{sec:model}

\subsection{Warm inflation}

The evolution of inflaton with a thermal friction in the Friedmann-Robertson-Walker metric is
\begin{align}
\ddot{\phi}+ (3H + \Gamma) \dot{\phi} + V_{,\phi}=0, \label{eq:phi}
\end{align}
where a dot denotes the derivative with respect to time, $V_{,\phi}=\partial V/\partial \phi$, and $\Gamma$ is the dissipative coefficient. From energy conservation, one can get the evolution of the radiation energy density,
\begin{align}
\dot{\rho}_r + 4 H \rho_r = \Gamma \dot{\phi}^2, \label{eq:radiation}
\end{align}
where the right-handed side is the energy transferred from inflaton, which sources the radiation bath.

To fully decode the inflaton dynamics, we also need the Friedman equation for the background evolution
\begin{align}
H^2 \simeq \frac{1}{3 M_{\rm pl}^2} \left( \frac{1}{2}\dot{\phi}^2 + V + \rho_r \right). \label{eq:Friedman}
\end{align}

In slow-roll regime, high order derivatives in Eq.(\ref{eq:phi}) and Eq.(\ref{eq:radiation}) can be neglected, i.e.,
\begin{align}
\ddot{\phi} \ll H \dot{\phi},\quad \dot{\rho}_r \ll H \rho_r. \label{eq:condition}
\end{align}

 As a result, the equation of motion of the inflaton reads
\begin{align}
\dot{\phi} \simeq - \frac{V_{,\phi}}{3 H + \Gamma} = - \frac{V_{,\phi}}{3 H (1+Q)}, \label{eq:phidot}
\end{align}
where we introduce the dimensionless dissipative ratio $Q\equiv\Gamma/(3H)$. The equation of motion of the radiation bath reads
\begin{align}
4 H \rho_r \simeq \Gamma \dot{\phi}^2. \label{eq:radeq}
\end{align}

During inflation, the inflaton potential energy dominates the energy density of the Universe, so the background evolution reads
\begin{align}
H^2 \simeq \frac{V}{3 M_{\rm pl}^2}. \label{eq:Hsq}
\end{align}

The slow-roll regime are parameterized by~\cite{Liddle:1994dx}
\begin{align}
\epsilon_w &\equiv \frac{\epsilon_V}{1+Q} = \frac{M_{\rm pl}^2}{2(1+Q)} \left( \frac{V_{,\phi}}{V} \right)^2;\\
\eta_w &\equiv \frac{\eta_V}{1+Q}=\frac{M_{\rm pl}^2}{ (1+Q)} \left( \frac{V_{,\phi\phi}}{V} \right);\\
\beta_w &\equiv \frac{M_{\rm pl}^2}{(1+Q)} \left( \frac{\Gamma_{,\phi} V_{,\phi}}{\Gamma V} \right).
\end{align}

To satisfy the requirements in Eq.(\ref{eq:condition}), by differentiating Eq.(\ref{eq:phidot}), Eq.(\ref{eq:radeq}) and Eq.(\ref{eq:Hsq}) one has~\cite{Taylor:2000ze,Hall:2003zp,Moss:2008yb,Visinelli:2011jy,Visinelli:2016rhn}
\begin{align}
\frac{\dot{H}}{H^2} &= -\epsilon_w, \\
\frac{\ddot{\phi}}{H \dot{\phi}} &= - \left( \eta_w - \beta_w + \frac{\beta_w - \epsilon_w}{1+Q} \right),\\
\frac{\dot{\rho}_r}{H \rho_r} &= - \left( 2\eta_w -\beta_w -\epsilon_w + 2 \frac{\beta_w - \epsilon_w}{1+Q} \right).
\end{align}

Being in the slow-roll regime requires that $\epsilon_w \ll 1, \eta_w \ll 1, \beta_w \ll 1$. With strong dissipative effects, i.e., $Q \gg 1$, we see that the slow-roll conditions can be satisfied with ``steep" potential, thus may help to have inflation in such otherwise not-possible inflation potentials. In general, including the dissipative term prolongs inflation, as can be easily seen.

The inclusion of the thermal dissipative term also alters the primordial scalar power spectrum. In cold inflation, quantum fluctuation of the inflaton field is the only source of the power spectrum. While in warm inflation, when $T>H$, the thermal fluctuation can dominate over the quantum fluctuation. With the temperature dependence of the dissipative coefficient, the inflaton and radiation fluctuation are coupled. Moreover, the inflaton can be excited and have a Bose-Einstein distribution rather than vacuum phase space distribution. The dimensionless primordial curvature power spectrum with all these effects is~\cite{Hall:2003zp,Graham:2009bf,BasteroGil:2011xd,Ramos:2013nsa,Benetti:2016jhf}
\begin{align}
\Delta_\mathcal{R}^2= \left( \frac{H^2}{2\pi \dot{\phi}} \right)^2 \left( 1+ 2 n_{\rm BE} + \frac{2\sqrt{3} \pi Q}{\sqrt{3+4\pi Q}}  \frac{T}{H} \right) G(Q), \label{eq:delR}
\end{align} 
where the quantities are all evaluated at horizon crossing, $n_{\rm BE}= 1/[{\rm exp}(H/T)-1]$ is the Bose-Einstein distribution function. The first factor is the same as that in cold inflation. The second and the third factors account for the thermal effects which arise due to the interaction with the thermal bath. The third factor accounts for the growth of inflaton fluctuation due to the coupling to radiation bath and can be get numerically.
 In $T \rightarrow 0, Q\rightarrow0$ limit, Eq.(\ref{eq:delR}) returns to the cold inflation power spectrum as expected. It is worth mentioning that this formula is neither confined to $T>H$ (thermal fluctuation dominant) nor $Q \gg 1$ (strong dissipative) cases. It actually allows for a ``smooth" transition between the cold and warm era. 
 
The spectral index can be evaluated as 
\begin{align}
n_s - 1 = \frac{d{\rm ln} \Delta_\mathcal{R}^2}{d{\rm ln}k} \simeq \frac{d{\rm ln} \Delta_\mathcal{R}^2}{dN}. \label{eq:ns1}
\end{align} 

The tensor fluctuation is approximately the same as in the cold inflation~\cite{Moss:2008yb}, and so is the tensor power spectrum
\begin{align}
\Delta_t^2 = \frac{2H^2}{\pi^2 M_{\rm pl}^2}.
\end{align}
Then one can determine the tensor-to-scalar ratio $r$ as
\begin{align}
r=\frac{\Delta_t^2}{\Delta_\mathcal{R}^2}. \label{eq:r1}
\end{align}

With the growth factor in $\Delta_\mathcal{R}^2$, one can see that warm inflation generally suppresses the tensor-to-scalar ratio compared to the cold case as $\Delta_t^2$ remains unchanged. If one recalls the natural inflation's predictions on $(n_s,r)$, suppressing $r$ will hopefully bring natural inflation into the Planck-allowed region, which we show in a subsequent section. 

\subsection{Warm natural inflation}

We consider the following potential for natural inflation
\begin{align}
V(\phi) = \Lambda^4 \left( 1+ \cos \phi/f \right).~\label{eq:V}
\end{align}

%where we admit the relation $\Lambda =\sqrt{m_\phi f}$ for an axion-like particle and choose to use $m_\phi$ and $f$ as variables.

Inspired from Ref~\cite{Berghaus:2019whh}, we assume inflaton couples to light gauge fields and experiences thermal friction with a coefficient of the following form
\begin{align}
\Gamma(T) = \kappa \alpha^5 \frac{T^3}{f_1^2}, \label{eq:gamma}
\end{align}
where $\alpha = \frac{g^2}{4 \pi}$ with $g$ being the gauge coupling, and $\kappa$ is a numerical factor which has a weak dependence on underlying gauge group color and flavor numbers. This friction originates in the inflaton's axion-like coupling to gauge fields, i.e.,
\begin{align}
\mathcal{L}=\frac{\alpha}{16\pi} \frac{\phi}{f_1} \tilde{G}_a^{\mu \nu} G_{\mu \nu}^a,\label{eq:L}
\end{align}
where $G_{\mu \nu}^a$ is a Yang-Mills gauge group field strength tensor. Protected by shift symmetry, inflaton potential does not suffer large thermal back-reactions associated with this friction. Here we use a different symbol $f_1$ because as pointed in Ref.~\cite{Berghaus:2019whh}, the UV potential is not necessarily in the same form as the IR potential. At high temperature, the sphaleron process is efficient and leads to topological charge fluctuation, which is responsible for the friction that inflaton feels. The friction coefficient is estimated from the sphaleron rate which is calculated in lattice gauge theory~\cite{Berghaus:2019whh,Moore:2010jd,Laine:2016hma}. The estimation is known to be valid when $m_{\phi} < \alpha^2 T$ and $H<\alpha^2 T$, where $m_\phi$ is the inflaton mass. Note that a dissipative coefficient with a cubic temperature dependence has been derived in Ref.~\cite{BasteroGil:2011xd,Bastero-Gil:2014raa} using a two-stage interaction configuration proposed in Ref.\cite{Berera:2001gs}.

Now we readily have
\begin{align}
\epsilon_w
&= \frac{1}{2(1+Q)}  \frac{M_{\rm pl}^2}{f^2} \frac{\sin^2 \phi/f}{(1+\cos\phi/f)^2};\\
\eta_w
&= - \frac{1}{(1+Q)} \frac{M_{\rm pl}^2}{f^2} \frac{\cos\phi/f}{1+\cos\phi/f};\\
\beta_w
&= \epsilon_w  \left(1-2f \frac{Q_{,\phi}}{Q} \frac{1+\cos\phi/f}{\sin\phi/f} \right).
\end{align}

The slow-roll parameter that violates the slow-roll conditions first gives us the field value at the end of inflation. Once we determined $\phi_{\rm end}$, we can get the field value $\phi_*$ at horizon crossing by requiring that by the end of inflation, we have 40-60 e-folds through
\begin{align}
N &= \int_{\phi_{\rm end}}^{\phi_{*} } \frac{1+Q}{M_{\rm pl}^2} \frac{V}{V_{,\phi}} d\phi \nonumber\\
&= \int_{\phi_*}^{\phi_{\rm end}} \frac{1+Q}{M_{\rm pl}^2} f \frac{1+\cos\phi/f}{\sin\phi/f} d\phi.
\end{align}
With the field value at the horizon crossing, one can work out the model's predictions like spectral index and tensor-to-scalar ratio according to Eq.(\ref{eq:ns}) and Eq.(\ref{eq:r}).

The growth factor in Eq.(\ref{eq:delR}) can be obtained numerically and in our case is given as~\cite{Benetti:2016jhf}
\begin{align}
G(Q) = 1+ 4.981 Q^{1.946} + 0.127 Q^{4.330}.
\end{align}

Putting together all the pieces in Eq.(\ref{eq:ns1}) and Eq.(\ref{eq:r1}), we get
\begin{align}
n_s = & 1- 6 \epsilon_w + 2 \eta_w + \frac{4Q}{1+7Q} (5\epsilon_w - 3\eta_w) \left( 1+\frac{1+Q}{2G(Q)} (9.693 Q^{0.946}+0.550 Q^{3.330})  \right) \nonumber\\
& +\frac{4}{1+7Q} \left( 1+2 n_{\rm BE} +\frac{2\sqrt{3} \pi Q}{\sqrt{3+4\pi Q} }\right)^{-1} \left[ n_{\rm BE}^2 \frac{H}{T} e^{H/T} \left[ 2(1+2Q) \epsilon_w -(1+Q) \eta_w \right] \right. \nonumber\\
&\left. + \frac{\sqrt{3} \pi Q T/H}{(3+4\pi Q)^{3/2} }  \left[ \epsilon_w \left( 21+9(3+2\pi)Q+26 \pi Q^2\right) -2\eta_w \left( 6+ (6+5\pi)Q +5\pi Q^2\right) \right]\right], \label{eq:ns}\\
r =& \frac{16\epsilon_w}{1+Q} \left( 1+2 n_{\rm BE} +\frac{2\sqrt{3} \pi Q}{\sqrt{3+4\pi Q} } \frac{T}{H} \right)^{-1}  \frac{1}{G(Q)},\label{eq:r}
\end{align}
where 
\begin{align}
\frac{T}{H} &=\left( \frac{9}{4 \tilde{g}_*} \frac{Q}{(1+Q)^2}  \frac{M_{\rm pl}^6 V_{,\phi}^2}{V^3}  \right)^{1/4} \nonumber\\
&=\left( \frac{9}{4 \tilde{g}_*} \frac{Q}{(1+Q)^2}  \frac{M_{\rm pl}^6 }{f^2 \Lambda^4} \frac{\sin^2 \phi/f}{(1+\cos\phi/f)^3}  \right)^{1/4}.
\end{align}

In the limit $Q\rightarrow 0, T\rightarrow 0$, Eq.(\ref{eq:ns}) and Eq.(\ref{eq:r}) recover the cold inflation result, i.e., 
\begin{align}
n_s &= 1- 6 \epsilon_V + 2 \eta_V,\\
r&= 16\epsilon_V.
\end{align}

\section{Results and discussions}\label{sec:result}

\subsection{Towards the Planck 2018 result: $f=5 M_{\rm pl}$}\label{sec:f5}

Cold natural inflation model's prediction in $(n_s,r)$ plane has an overlap with the $2\sigma$ region of the Planck 2018 result, but none overlap in $1\sigma$ region. In this section, we will show that with the inclusion of the thermal friction, natural inflation's prediction can stay inside the $1\sigma$ region of the Planck 2018 result.

Given the temperature dependence in the friction coefficient, we can work out the temperature dependence of $Q(T)$ through Eq.(\ref{eq:phidot}), Eq.(\ref{eq:Hsq}), Eq.(\ref{eq:gamma}) and Eq.(\ref{eq:radeq}). Note also $\rho_r=\frac{\pi^2}{30} g_* T^4 \equiv\tilde{g}_* T^4$, where $g_*$ is the number of relativistic degrees of freedom in the thermal bath. We find
\begin{align}
Q (1+Q)^6 &= \frac{ (\kappa \alpha^5)^4}{576 \tilde{g}_*^3}  \frac{M_{\rm pl}^{10} \Lambda^4 }{f^6f_1^8}   \frac{(\sin \phi/f)^6}{(1+\cos \phi/f)^5}\nonumber \\
&= \frac{c}{\tilde{f}^6}   \frac{(\sin \phi/f)^6}{(1+\cos \phi/f)^5}, \label{eq:Q}
\end{align}
where in the last line we use $\tilde{f}=f/M_{\rm pl}$ and introduce the dimensionless parameter
\begin{align}
c&= \frac{(\kappa \alpha^5)^4}{576 \tilde{g}_*^3} \frac{M_{\rm pl}^4 \Lambda^4}{f_1^8} \nonumber \\
&= 4.8 \times 10^{-20} \left(\frac{\kappa \alpha^5}{10^{-3}} \right)^4 \left( \frac{\tilde{g}_*}{33} \right)^{-3} \frac{M_{\rm pl}^4 \Lambda^4}{f_1^8},
\label{eq:c}
\end{align}
where in the last row we assume for an illustrative purpose the standard model degrees of freedom, i.e., $g_*\simeq 100$, and $\kappa \simeq 100, ~\alpha \simeq 0.1$. A unique positive value of $Q$ can be get since the right handed side of Eq.(\ref{eq:Q}) is positive.
With no a priori knowledge on $\Lambda$ and $f_1$, $c$ can span a large range. It is argued that the scale of infrared potential should be no larger than that of the ultraviolet potential~\cite{Goswami:2019ehb}, namely, $f_1\lesssim \Lambda$. In this case we get a rather loose lower bound on $c$: $c \gtrsim 4.8 \times 10^{-12}$. For a fixed $f$, the parameter $c$ characterizes the magnitude of the dissipative effect as it determines $Q$ through Eq.(\ref{eq:Q}). We do not assume a strong dissipative effect, as the natural inflation already has a flat potential.  

In our model, all the slow-roll parameters as well as the dissipative ratio $Q$ can be viewed as functions of dimensionless parameters $\tilde{\phi},~$c$,~\tilde{f}$, where $\tilde{\phi}=\phi/M_{\rm pl}, \tilde{f}=f/M_{\rm pl}$. It is understood that $\tilde{\phi}$ and $\tilde{f}$ are the field value and the decay constant written in $M_{\rm pl}$ unit.

Since the observed scalar spectral tilt favors a large $f$ for the natural inflation model, we start with a relatively large $f$. We adopt $f= 5 M_{\rm pl}$ as a benchmark point to show the effect of thermal friction in our considered natural warm inflation model.

\begin{figure}[h!]
\centering
\includegraphics[width=1.\textwidth]{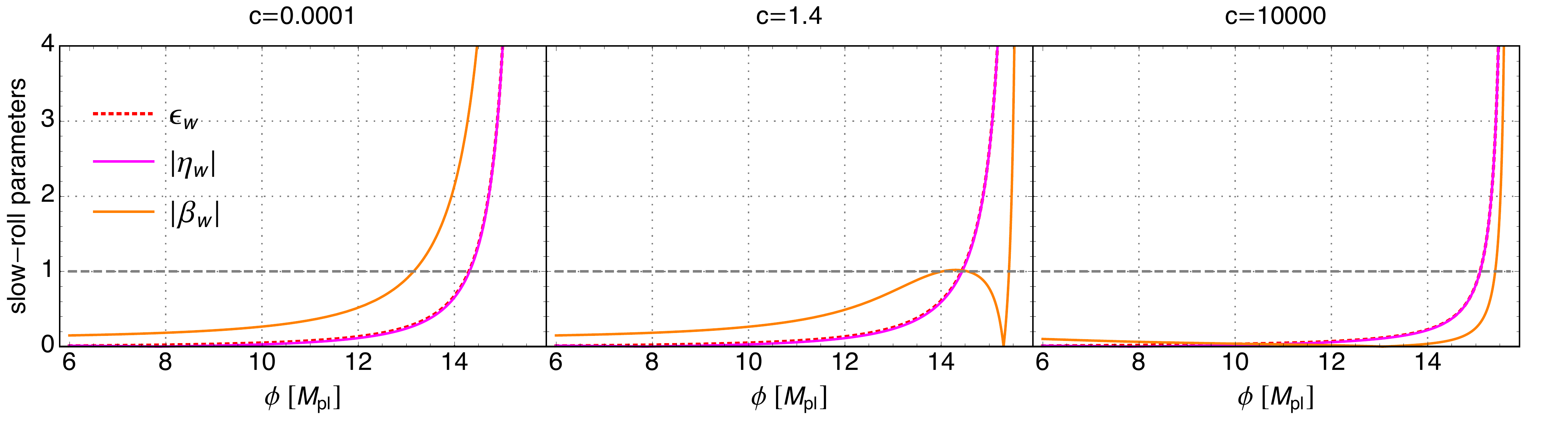}
\includegraphics[width=0.49\textwidth]{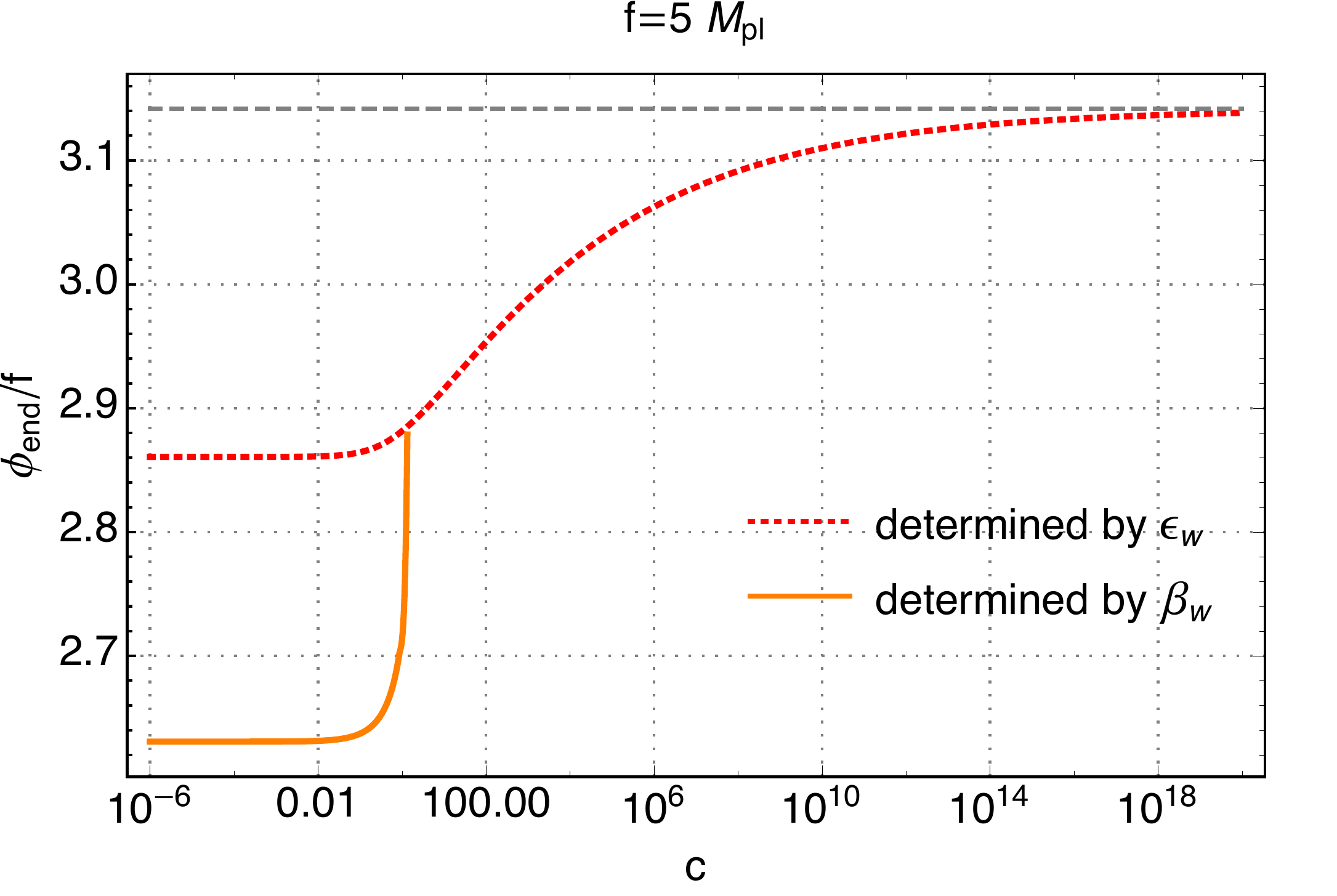}
\includegraphics[width=0.49\textwidth]{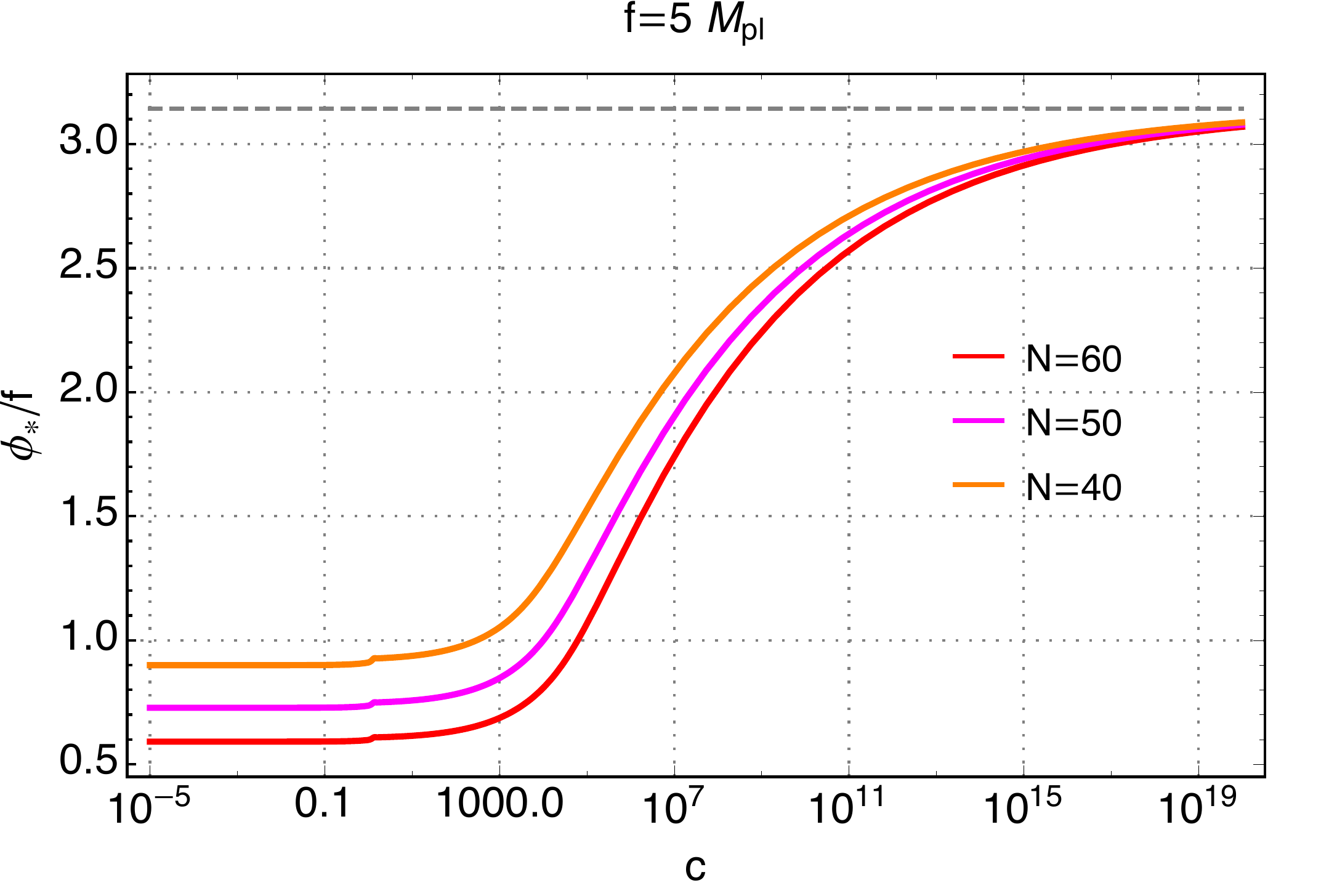}
\caption{Results for $f=5 M_\mathrm{pl}$. First row: the slow-roll parameters with benchmark points of $c$. For $c\lesssim 1.4$, $\beta_w$ violates the slow-roll conditions first, while for $c>1.4$, $\epsilon_w$ violates the slow-roll conditions first. Second row: field value at the end of inflation (left) and at the horizon crossing (right) as a function of the dimensionless parameter $c$. Results in these plots do not depend on inflaton thermalization and are valid in both cases.}
\label{fig:f5_1}
\end{figure}

\begin{figure}[h!]
\centering
\includegraphics[width=0.49\textwidth]{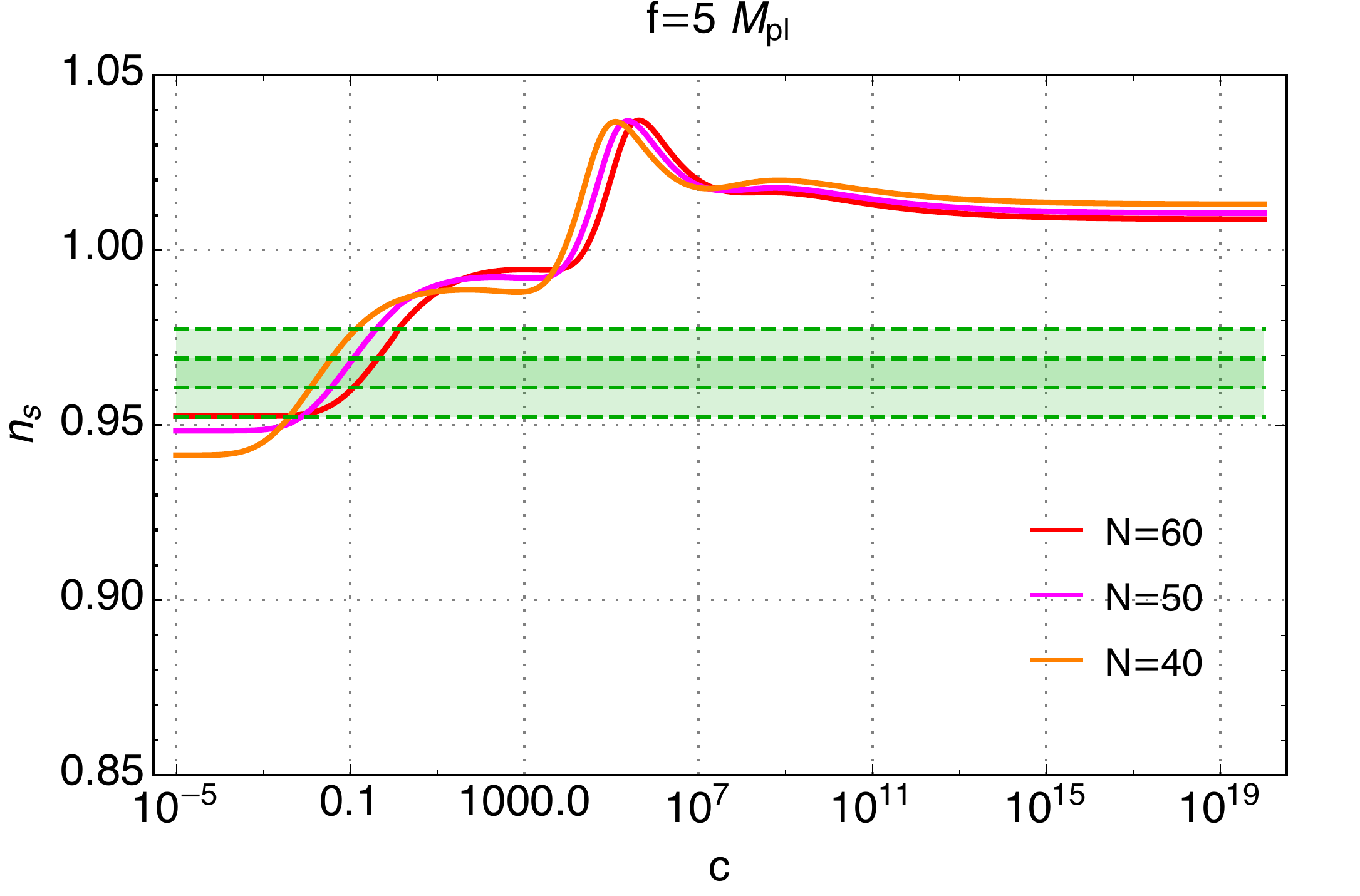}
\includegraphics[width=0.49\textwidth]{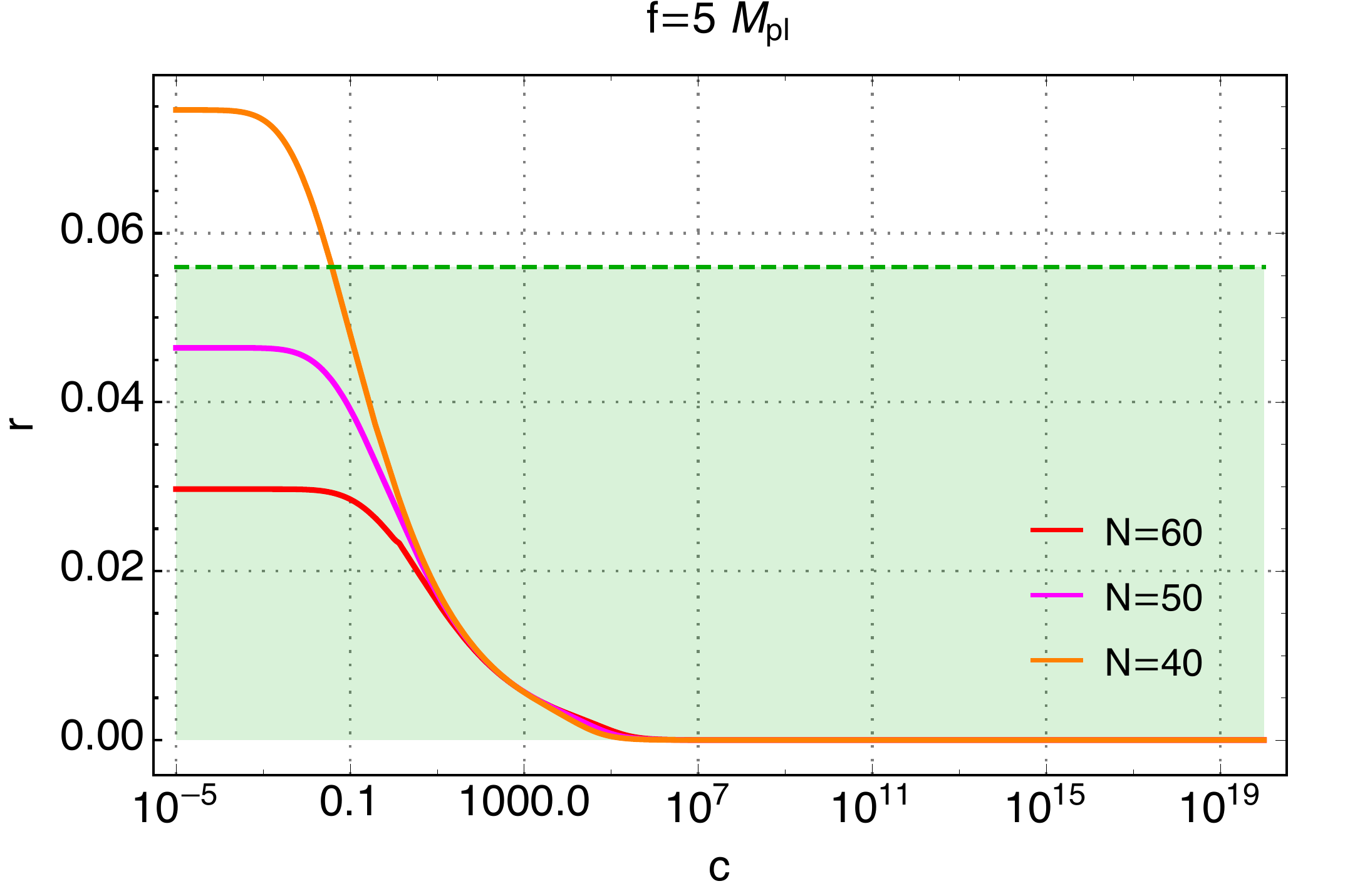}
\includegraphics[width=0.48\textwidth]{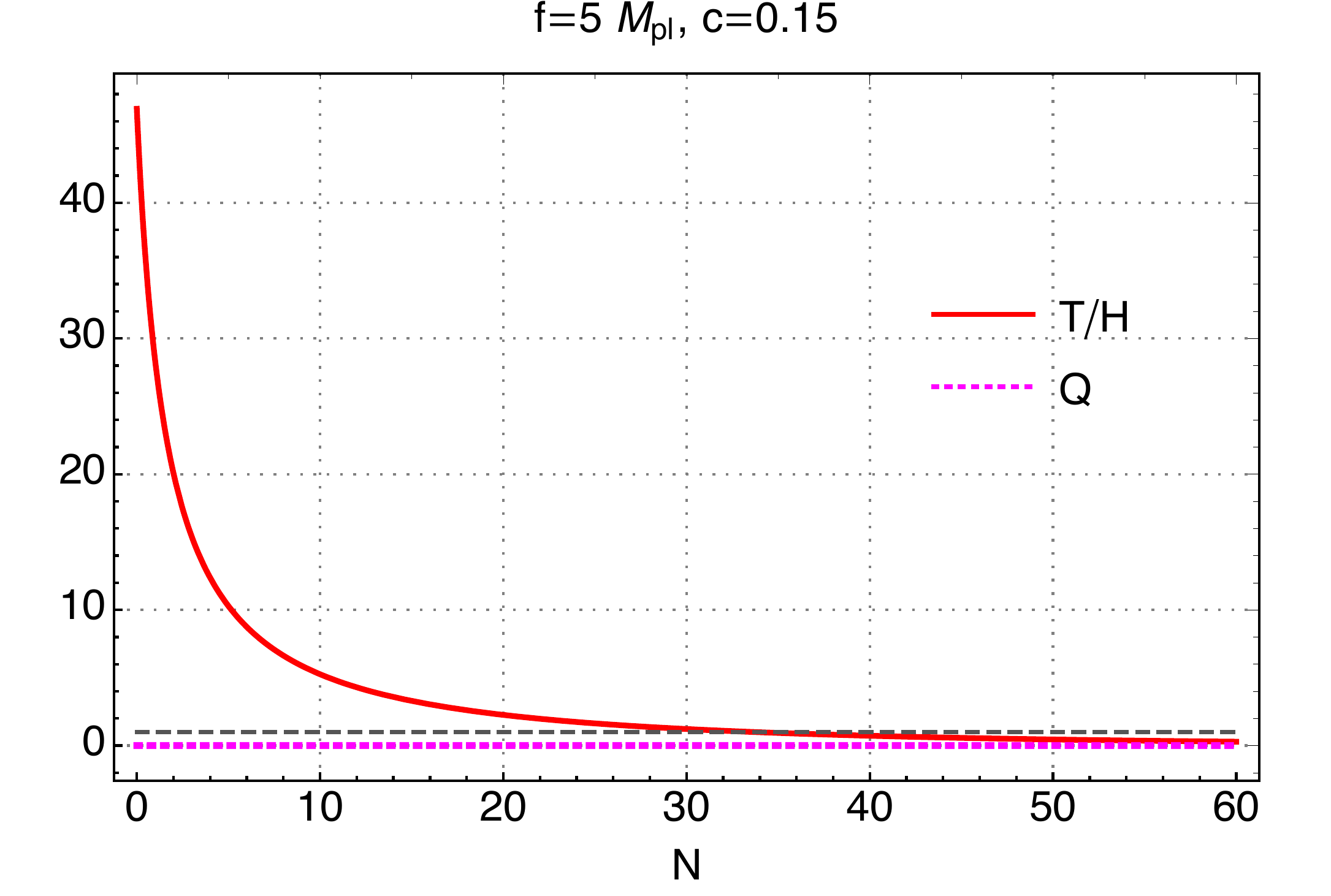}
\includegraphics[width=0.5\textwidth]{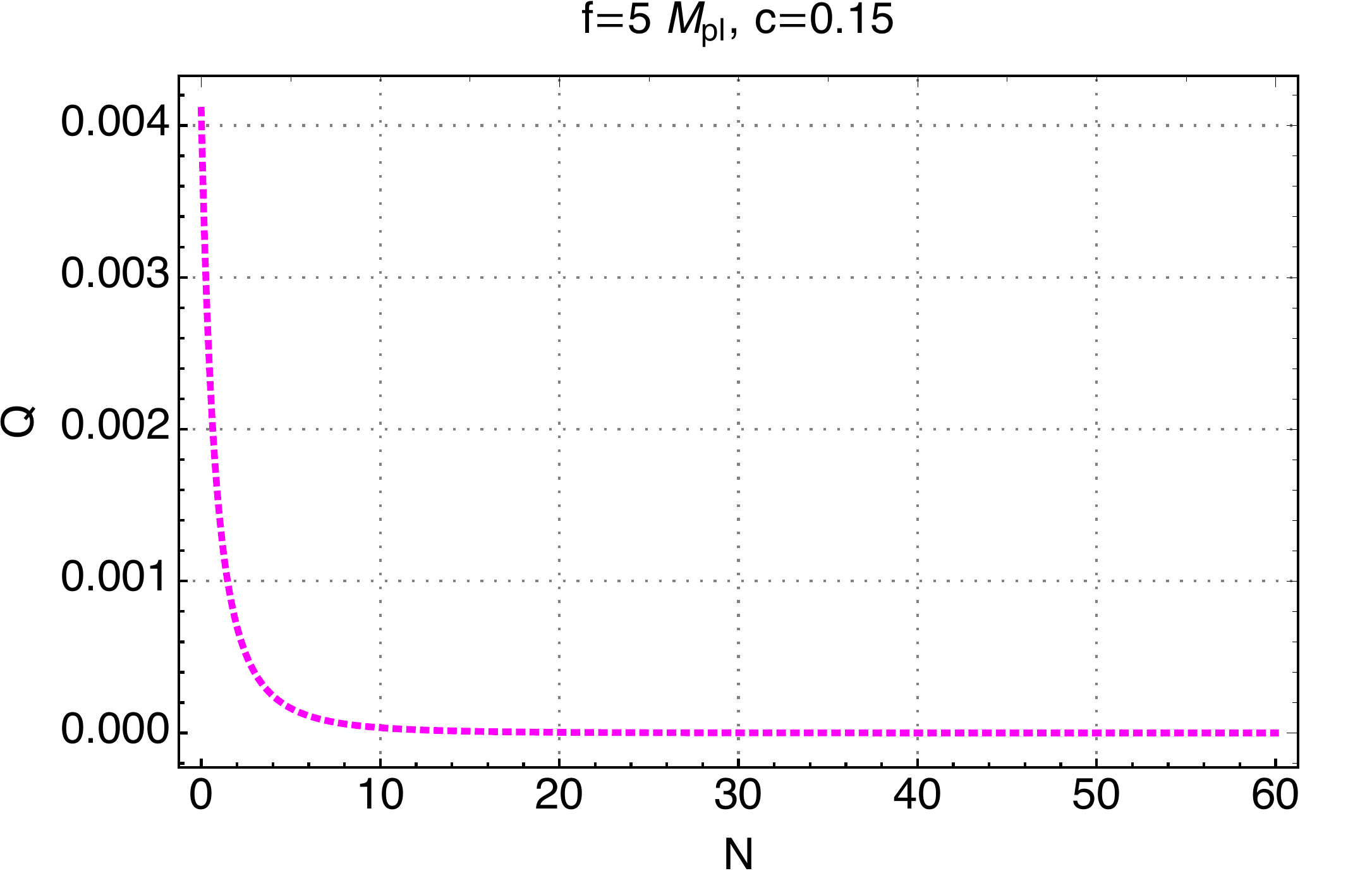}
\caption{Results for $f=5 M_{\mathrm{pl}}$ in thermalized inflaton case. First row: scalar spectral index (left) and the tensor-to-scalar ratio (right) as a function of the parameter $c$. The green bands are allowed regions given by the Planck 2018 results~\cite{Akrami:2018odb}. Second row: the evolution of the $T/H$ ratio (red line) and the dissipative ratio Q (magenta dotted line). On the right plot, we zoom in on the Q evolution. The horizontal axis is the number of e-folds before the end of inflation, so the evolution starts from the right (e.g., $N=60$) and evolves to the left ($N=0$).
}
\label{fig:f5_2}
\end{figure}

\begin{figure}[h]
\centering
\includegraphics[width=.7\textwidth]{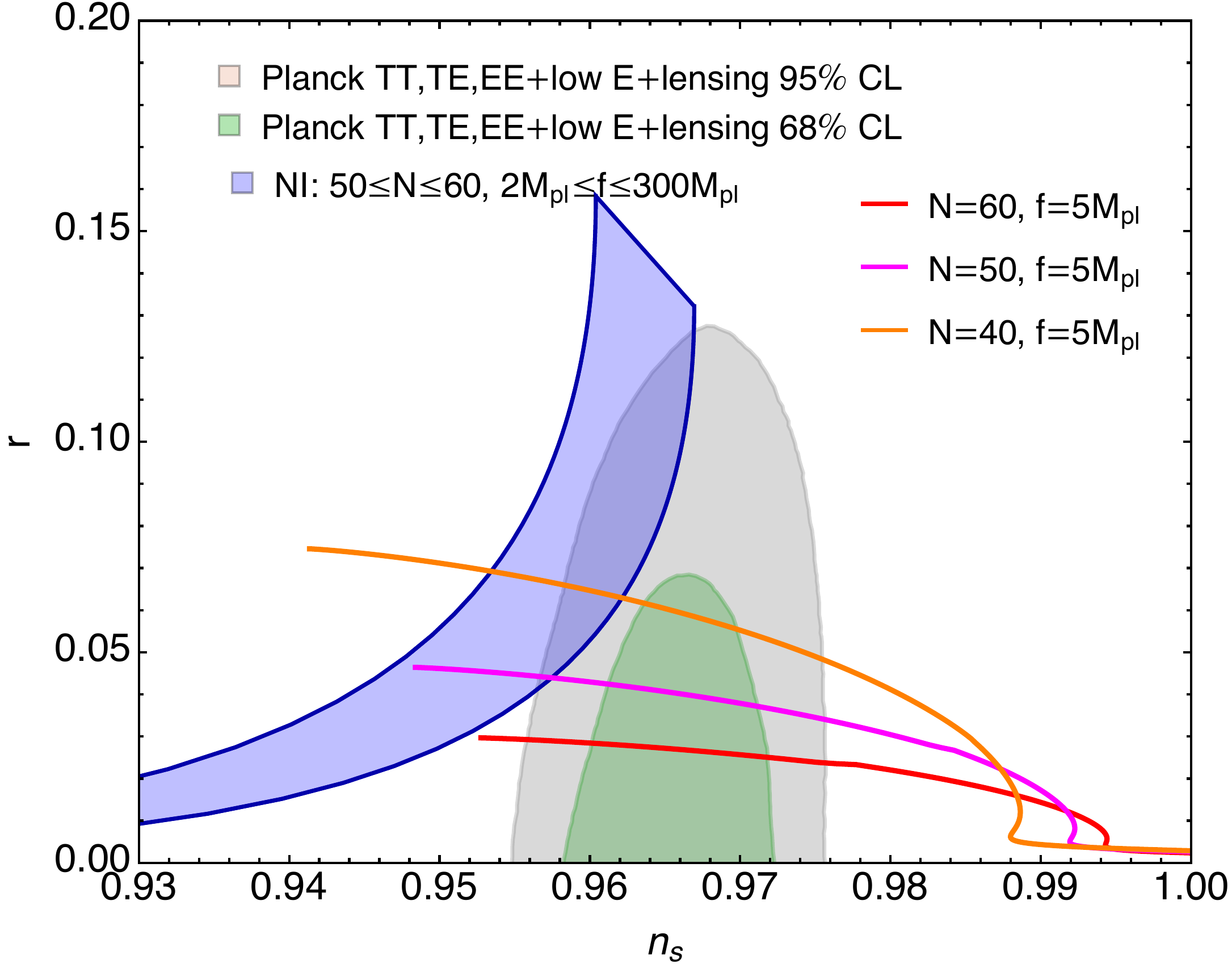}
\caption{ $n_s - r$ plot with Planck 2018 constraints~\cite{Akrami:2018odb}. We also show the cold natural inflation results (the blue region) for comparison. Here inflaton is thermalized.
}
\label{fig:f5_3}
\end{figure}

We show the numerical results in $f=5 M_{\mathrm{pl}}$ case in Fig.~\ref{fig:f5_1}, Fig.~\ref{fig:f5_2} and Fig.~\ref{fig:f5_3}. In the first row of Fig.~\ref{fig:f5_1}, we show slow-roll parameters with different values of $c$. When $c\simeq 1.4$, we get $\epsilon_w \simeq |\eta_w| \simeq |\beta_w| \simeq 1$ at a same field value. For $c\lesssim 1.4$, $\beta_w$ violates the slow-roll conditions first, while $c>1.4$, $\epsilon_w (\eta_w)$ violates the slow-roll conditions first. So we use $|\beta (\phi_{\rm end})| = 1$ to determine the field value at the end of inflation when $c\lesssim 1.4$, and use $\epsilon_w (\phi_{\rm end}) = 1$ otherwise. We show the field value at the end of inflation on the left plot in the second row. On the right plot, we show the field value at the horizon crossing for $N=40, 50, 60$. 

With this field value, we can determine the scalar spectral index and tensor-to-scalar ratio according to Eq.(\ref{eq:ns}) and Eq.(\ref{eq:r}), we plot the scalar spectral index and the tensor-to-scalar ratio as a function of the parameter $c$ in the first row of Fig.~\ref{fig:f5_2}, where the green bands are the allowed regions from the Planck 2018 results. We see that the Planck results prefer a small $c$, which is $[0.013,1.2]$ for $N=60$, which leads to $\left(\frac{M_{\rm pl}}{f_1} \right)^8  \left( \frac{\Lambda}{M_{\rm pl}} \right)^4 \in [2.7\times 10^{17}, 2.5\times 10^{19}]$. For example, $c=1$ and $\Lambda=10^{16}$ GeV result in $f_1=6\times 10^{14} {\rm GeV}$. Also in the first row of Fig.~\ref{fig:f5_2}, we see that the inclusion of the thermal effects generally increase $n_s$ and suppress $r$. For very small $c$ (corresponds to very small $Q$), $n_s$ and $r$ nearly stay the same as expected. A noticeable change happens around $c \simeq 10^{-4}$, where $n_s$ starts growing and $r$ experience decreasing. $r$ continues decreasing for growing $c$ while $n_s$ first grows and then slightly decreases and is finally asymptotic to $1$. This can be seen from Eq.(\ref{eq:ns}) in the large $Q$ limit.

To further show the dynamics during inflation, we plot the $T/H$  and $Q$ evolution in the last row of Fig.~\ref{fig:f5_2}. Note that the number of e-folds counts from the end of inflation back to the horizon crossing, which means $N=0$ corresponds to the end of inflation and $N=60 (50, 40)$ at the horizon crossing. Looking from right to left, we see that both $T/H$ and $Q$ increase during inflation as expected. Inflation starts with $T/H < 1$, i.e., cold era, with the coupling to the light fields, it evolves to a $T/H > 1$, i.e., warm era. During the whole inflation period, we have $Q<1$, i.e., in the weak dissipative regime. $Q$ starts quickly growing near the end of inflation, which signals a starting of a reheating era.

We also show the $n_s - r$ plot with the Planck 2018 constraints~\cite{Akrami:2018odb} in Fig~\ref{fig:f5_3}. We see that there is parameter space compatible with the Planck 2018 results even at the $1\sigma$ level. In the vanishing $c$ limit, the lines go back to the cold natural inflation prediction as expected. Note that the cold natural inflation's predictions at $f=5M_{\rm pl}$ lie outside the $2\sigma$ region and have been excluded. So we successfully rescue the natural inflation at $f=5M_{\rm pl}$.

Now we have the whole picture of our model. Inflation starts when the Universe is cold. With a coupling to the light fields, the dissipative term quickly sources a thermal bath of radiation. Though the thermal equilibrium is established very soon, the equilibrium temperature is lower than $H$ for the first few tens of e-folds, and the inflation looks cold. The dissipative effect contributes to a thermal fluctuation that grows and eventually dominates over the quantum fluctuation of the inflaton field. The presence of the thermal dissipative effect, although weak, changes the predictions of natural inflation and brings it into an agreement with the Planck 2018 results.

\subsubsection{Dissipative coefficient from the axion-like interaction}

It is important to notice that our results obtained in the previous section are not confined to a specific origin of the dissipative coefficient. Although an axion-like interaction inspires our investigation of the cubic temperature-dependent dissipative coefficient, our derivation and results are valid as long as the dissipative coefficient has cubic temperature dependence, regardless of its microphysical origin. It is still of great interest to investigate the case that the dissipative coefficient comes from an axion-like interaction as shown in Eq.(\ref{eq:L}).

As is mentioned, the estimation of the dissipative coefficient from the axion-like interaction in Eq.(\ref{eq:L}) is only known to be valid when $m_{\phi} < \alpha^2 T$ and $H<\alpha^2 T$~\cite{Berghaus:2019whh}. Let us focus on the latter one for definiteness.  As the gauge group is not necessarily the standard model gauge group, $\alpha$ is only bounded from perturbativity. The lattice calculation of the sphaleron rates is valid for $\alpha \lesssim 0.1$~\cite{Moore:2010jd}. Look at the second row of Fig.~\ref{fig:f5_2} we see a tension with the requirement $H<\alpha^2 T$ even for $\alpha \simeq 0.1$.

The tension gets stronger if we consider extra $\alpha$ constraint from inflaton thermalization. Applying Eq.(\ref{eq:delR}) we do assume that inflaton is thermalized, which requires the interaction rate with the gauge fields $\Gamma_{\phi g}$ larger than the Hubble expansion rate. $\Gamma_{\phi g}$ is estimated from Eq.(\ref{eq:L}) as~\cite{Berghaus:2019whh}
\begin{align}
\Gamma_{\phi g} \simeq \frac{\alpha^3 T^3}{32\pi f_1^2} = \frac{1}{32\pi \kappa \alpha^2} \Gamma.
\label{eq:GammaPhi}
\end{align}
With $\Gamma= 3QH$, $\Gamma_{\phi g}>H$ requires $\frac{3Q}{32\pi \kappa \alpha^2}>1$, which is satisfied when $\alpha < 10^{-2} \sqrt{Q}$. With the Q value at the horizon crossing in our model, it roughly reads $\alpha < 10^{-5}$, which clearly violates the conditions $m_\phi < \alpha^2 T$ and $H<\alpha^2 T$. 

Several comments are in order. First, not satisfying the condition $m_{\phi} < \alpha^2 T$ and $H<\alpha^2 T$ means that we do not know from the thermal field theory calculation that the friction coefficient can be written in the form as Eq.(\ref{eq:gamma}). Neither does it rule out this possibility. Second, if future development in thermal field theory relaxes the requirement, we confront first the lattice limit. The limit $\alpha \lesssim 0.1$ set by current lattice calculation is more a technical reason than a physical one. Physically, $0.1 <\alpha< 1$ are allowed, and a larger $\alpha$ alleviates the tension. Third, the inflaton is not necessarily thermalized with the plasma. When it is the case, $\alpha$ does not receive extra constraint and is only bounded from perturbativity. We discuss this case in the subsequent subsection in detail. Last but not least, this tension exists only when we assume the axion-like interaction to be the origin of the cubic temperature-dependent dissipative coefficient. Our results are valid as long as the dissipative coefficient has cubic temperature dependence, regardless of its origin. 

%Actually, our calculation is mostly based on the dimensionless parameter $c$, which may or may not be in the form as shown in Eq.(\ref{eq:c}). What is essential to us is that the friction coefficient has temperature dependence as $\Gamma \propto  T^3$.

%Now we have the whole picture of our model. Inflation starts when the Universe is cold. With a coupling to the light fields, the dissipative term quickly sources a thermal bath of radiation. Though the thermal equilibrium is established very soon, the equilibrium temperature is lower than H for the first few tens of e-folds, and the inflation looks cold. The dissipative effect contributes a thermal fluctuation which grows and eventually dominates over the quantum fluctuation of the inflaton field. The presence of the thermal dissipative effect, although weak, changes the predictions of natural inflation and brings it into agreement with Planck 2018 result.

\subsubsection{Non-thermalized inflaton}

Let us consider the case that inflaton is not thermalized with the plasma. The only modification is to replace the thermal distribution function of the inflaton in Eq.(\ref{eq:ns}) and Eq.(\ref{eq:r}) to its out-of-equilibrium distribution, which is estimated as~\cite{Bastero-Gil:2017yzb}
\begin{align}
n \simeq \frac{\gamma_{\phi g}}{3 + \gamma_{\phi g}} n_\mathrm{BE},
\end{align}
where $\gamma_{\phi g} \equiv \Gamma_{\phi g}/H$ and $\Gamma_{\phi g}$ is defined in Eq.(\ref{eq:GammaPhi}).

\begin{figure}[h]
\centering
\includegraphics[width=0.49\textwidth]{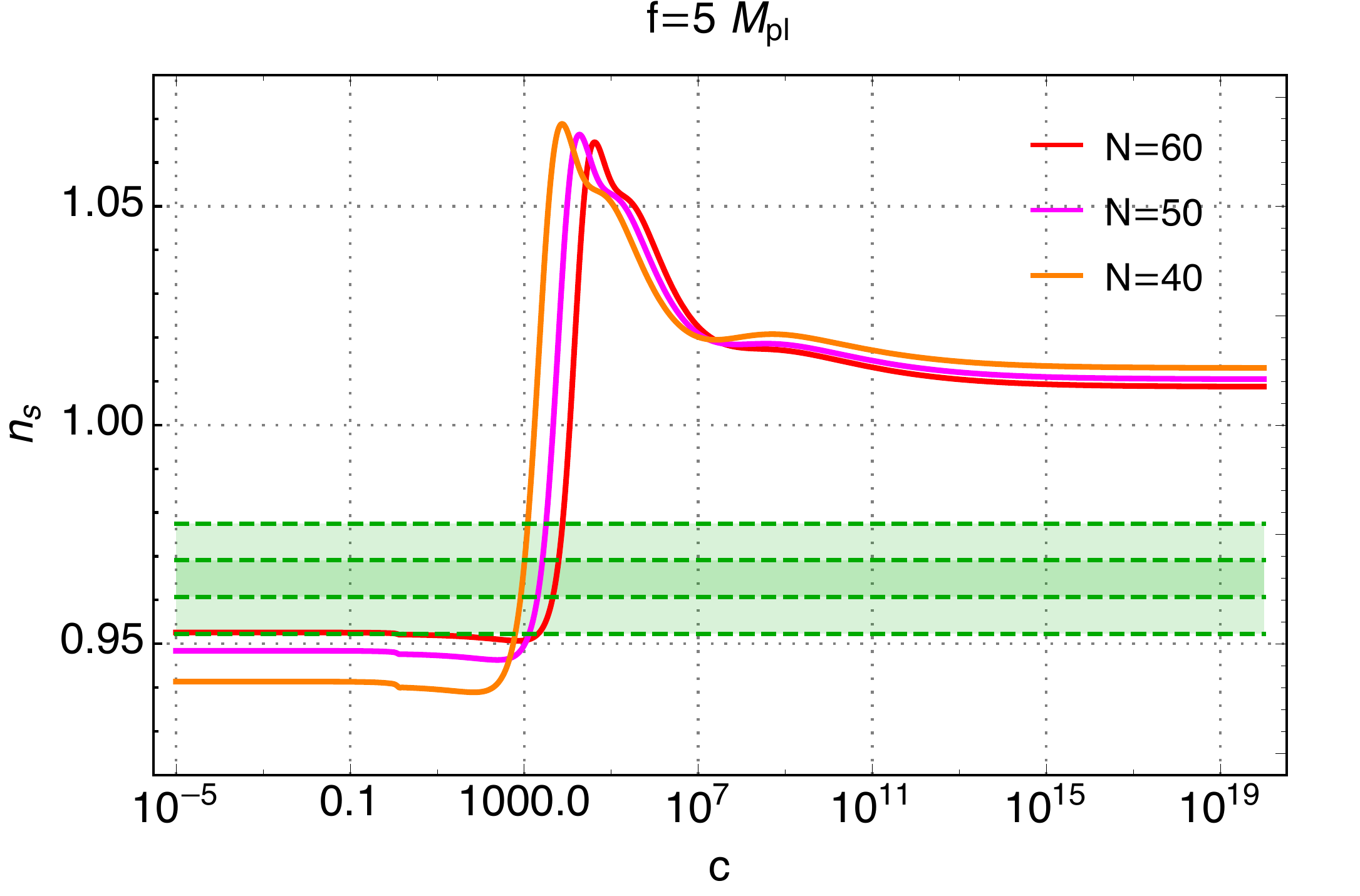}
\includegraphics[width=0.49\textwidth]{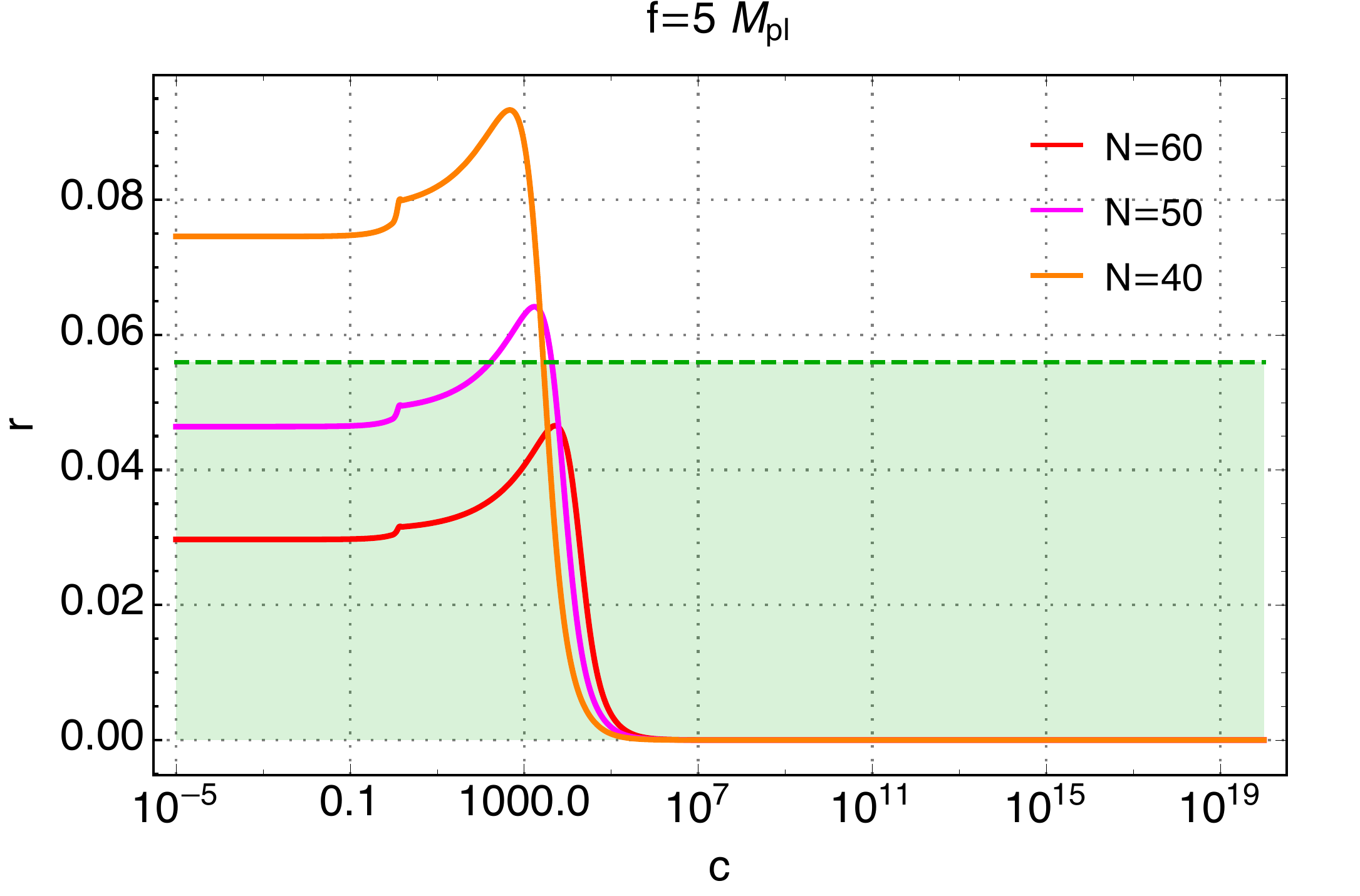}
\includegraphics[width=0.48\textwidth]{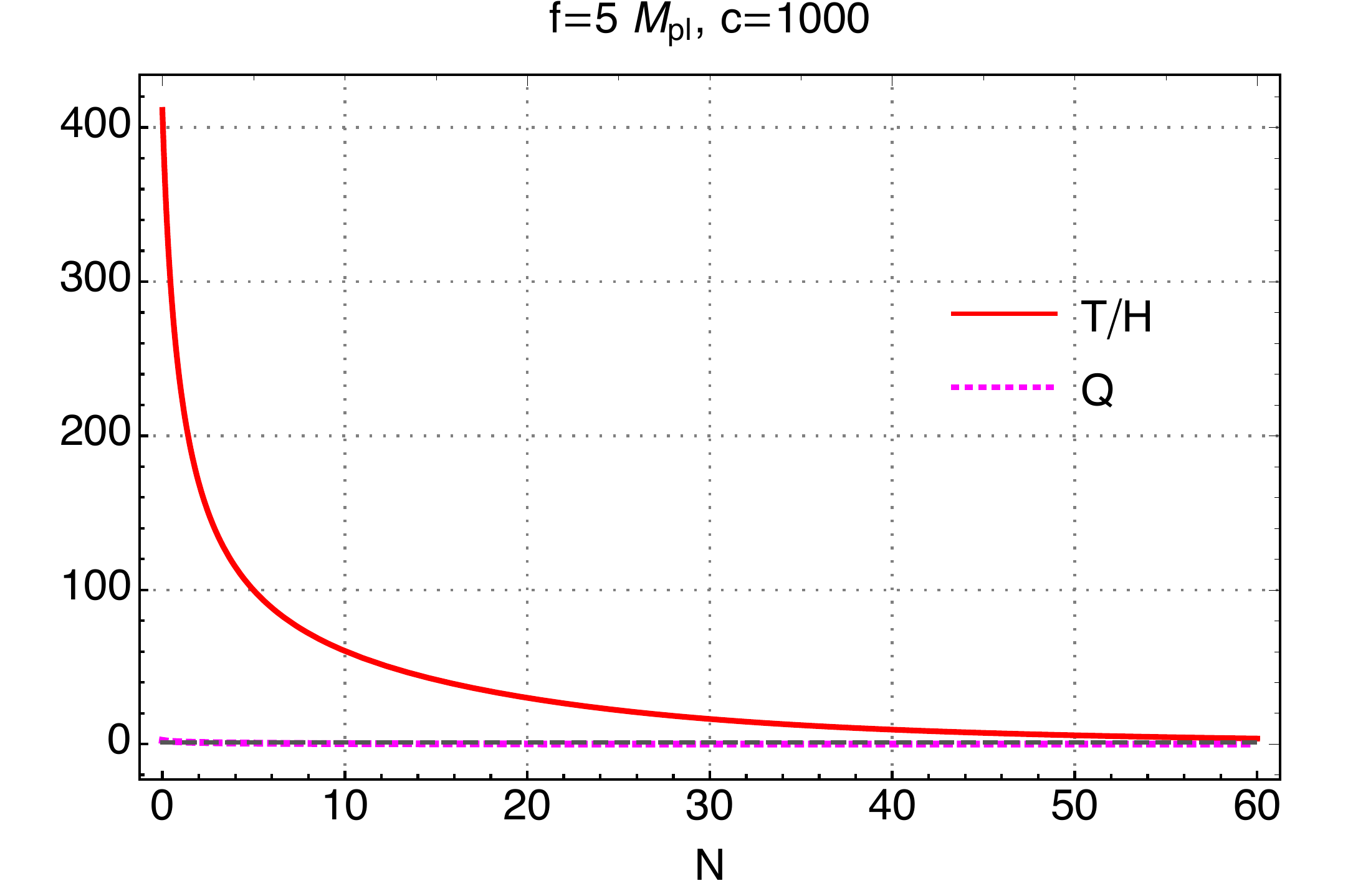}
\includegraphics[width=0.5\textwidth]{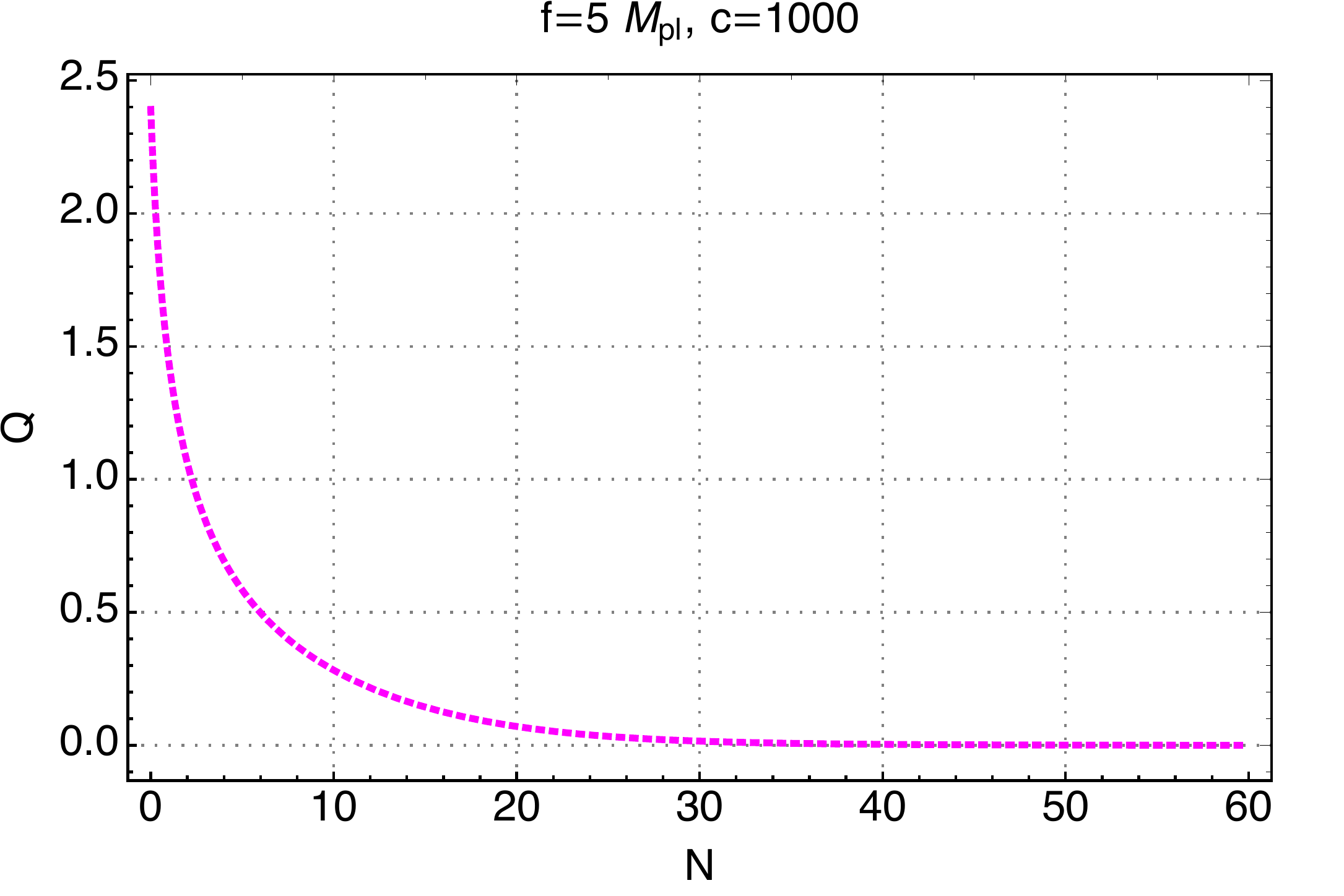}
\caption{Numerical results in a non-thermalized inflaton case with $f=5 M_\mathrm{pl}$. First row: scalar spectral index (left) and the tensor-to-scalar ratio (right) as a function of the parameter $c$. The green bands are allowed regions given by the Planck 2018 results~\cite{Akrami:2018odb}. Second row: the evolution of the $T/H$ ratio (red line) and the dissipative ratio $Q$ (magenta dotted line). On the right plot we zoom in on the $Q$ evolution. The horizontal axis is the number of e-folds before the end of inflation, so the evolution starts from the right (e.g., $N=60$) and evolves to the left ($N=0$).
}
\label{fig:nonthermal_1}
\end{figure}

\begin{figure}[h]
\centering
\includegraphics[width=.7\textwidth]{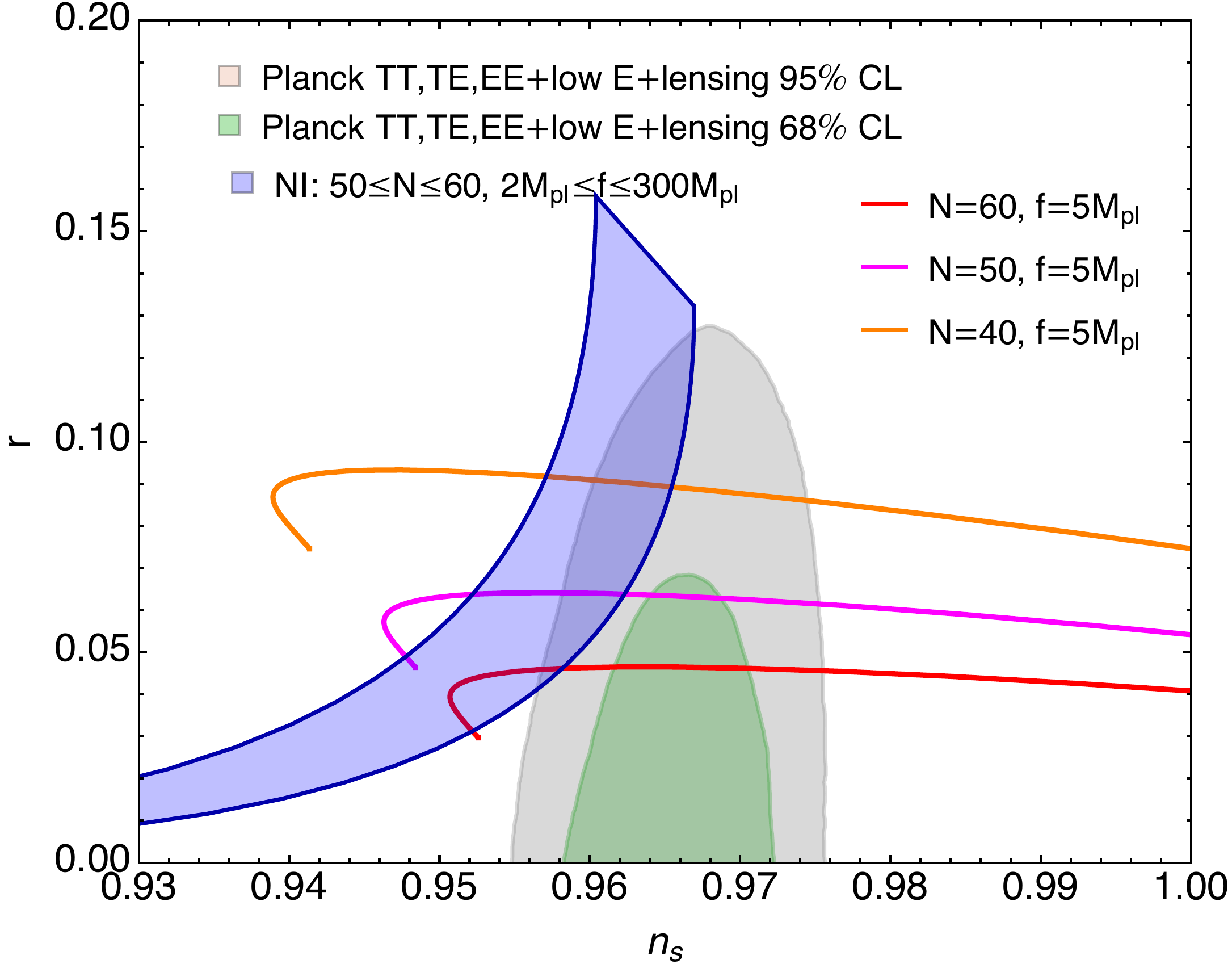}
\caption{ $n_s - r$ plot with the Planck 2018 constraints~\cite{Akrami:2018odb} in the case of non-thermalized inflaton. We also show the cold natural inflation results (the blue region) for comparison. The little excess of warm natural inflation lines over the cold natural inflation bands is explained in text.
}
\label{fig:nonthermal_2}
\end{figure}

Let us still look at the $f=5 M_{\mathrm{pl}}$ case so that we can make a direct comparison with the thermalized inflaton case. Note that the slow-roll parameters are unchanged, so are the field values at both the end of inflation and horizon crossing. We show the results for the scalar spectral index and the tensor-to-scalar ratio in the first row of Fig.~\ref{fig:nonthermal_1}. Compared with that of Fig.\ref{fig:f5_2}, we find the behaviors of $n_s$ and $r$ as functions of $c$ are qualitatively the same. For the viable regions, non thermalized inflaton case prefers a larger $c$ than before.

The evolutions of the $T/H$ ratio and the dissipative ratio $Q$ are shown in the second row of Fig.~\ref{fig:nonthermal_1}, where we choose a conservative value of $c$ to be $1000$ and use $\alpha=0.1$. A larger $c$ can further alleviate the tension. For our choice of $c$, we have $Q\simeq 0.0002,~T/H \simeq 3.6$ at the start of inflation and $Q\simeq 2.4,~T/H \simeq 411$ at the end for $N=60$. The non-thermalized inflaton slightly moves the warm inflation from a purely weak regime to an intermediate regime. It is intuitively understood as the saved energy from thermalizing the inflaton now heats the whole system.  

We show the results in $n_s - r$ plane in Fig.~\ref{fig:nonthermal_2}. We find dramatic differences compared with that of Fig.~\ref{fig:f5_3}. For a given $n_s$, $r$ is generally larger and $N=40$ line is outside the $1\sigma$ region now. One can also identify a suspicious excess of the warm natural inflation lines over the cold natural inflation bands: they intersect first, and the warm lines go a little further. The intersection happens at a $Q$ value at $\mathcal{O}(10^{-6})$, but the warm lines stop when $Q \simeq10^{-13}$. This can be understood as follows. If one admits $f_1\lesssim \Lambda$, there is a lower bound on $c$ as discussed in section~\ref{sec:f5}. As a result, $Q$ cannot be absolutely zero. When $c$ is small enough, which leads to a nearly vanishing $Q$, one expects the warm natural inflation lines to approach that of the cold natural inflation lines -- which is the generic behaviors as can be seen in Figs.~\ref{fig:f5_3} and \ref{fig:nonthermal_2}. As for the little excess, one can look back at the first row of Fig.~\ref{fig:f5_1}, where we show the slow-roll parameters as a function of the field with different $c$ values. We see that the end of inflation is determined by $\epsilon_w$ only if $c \gtrsim 1.4$. For $c \lesssim 1.4$, it is $\beta_w$ violates the slow-roll conditions first, which leads to a $\phi_\mathrm{end}$ different from that determined from $\epsilon_w$ as is shown in the lower-left plot of Fig.~\ref{fig:f5_1}. Remember that there is no $\beta$ can be defined in cold inflation, and that is why the little discrepancy in Fig.~\ref{fig:nonthermal_2} exists. Focus on $N=60$ lines for an example, it should be emphasized that both the intersections with the cold natural inflation lines do not correspond to $f=5 M_\mathrm{pl}$ for cold natural inflation, but the lower intersecting point ($f=5.00047 M_\mathrm{pl}$) is very close to that.  This fits our intuition: for a fixed $f$, if we allow $Q$ to decrease to zero, we can go back to cold natural inflation.

To sum up, if the inflaton is not thermalized, the tension with the thermal field theory requirement is much alleviated depending on the $\alpha$ value of the underlying theory but still not disappears. Non-thermalized inflaton helps to alleviate the tension with the thermal field theory requirement but fits worse to the observations than the thermalized inflaton case. So we stick to the thermalized inflaton case from now on.

\subsection{Towards a smaller decay constant $f$} 

As is mentioned, $f>1 M_{\rm pl}$ may cause difficulty in embedding into a more fundamental theory. For this reason, we want to find out whether $f< 1 M_{\rm pl}$ is viable in this model. Given the complexity in Eq.(\ref{eq:ns}) and Eq.(\ref{eq:r}), it is difficult to get an analytical understanding. Note that $Q$ is not necessarily large (as the $f=5M_{\rm pl}$ results show), so it is not possible to further simplify the expressions. Our efforts towards a smaller $f$ include two steps: First, we show the numerical result in $f=1M_{\rm pl}$ case, which is generic in $c$; Then we fix $c$ and find the spectral index's dependence on $f$. Based on these findings, we conclude on the validity of $f< 1 M_{\rm pl}$.

\begin{figure}
\centering
\includegraphics[width=0.49\textwidth]{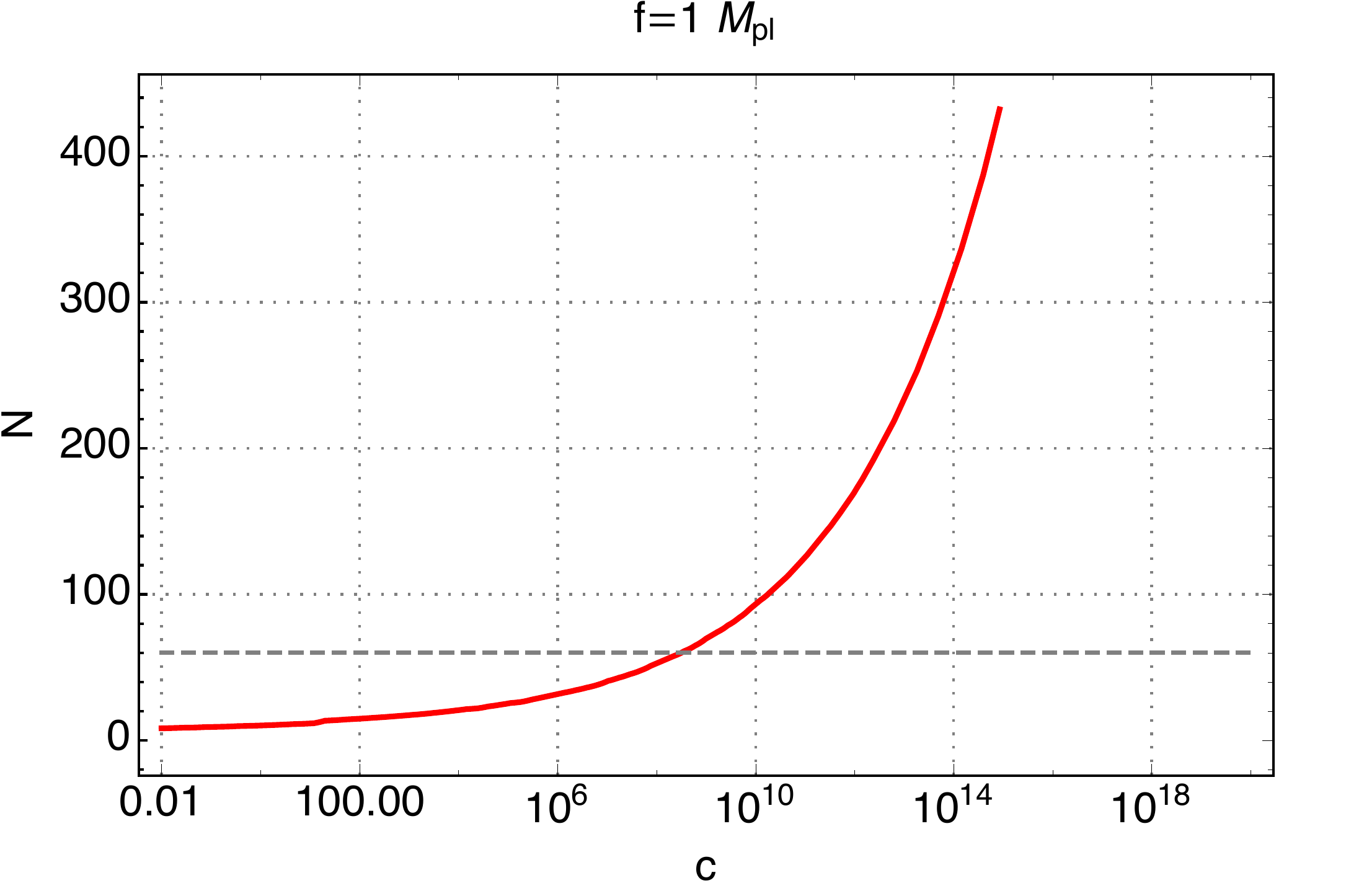}
\includegraphics[width=0.49\textwidth]{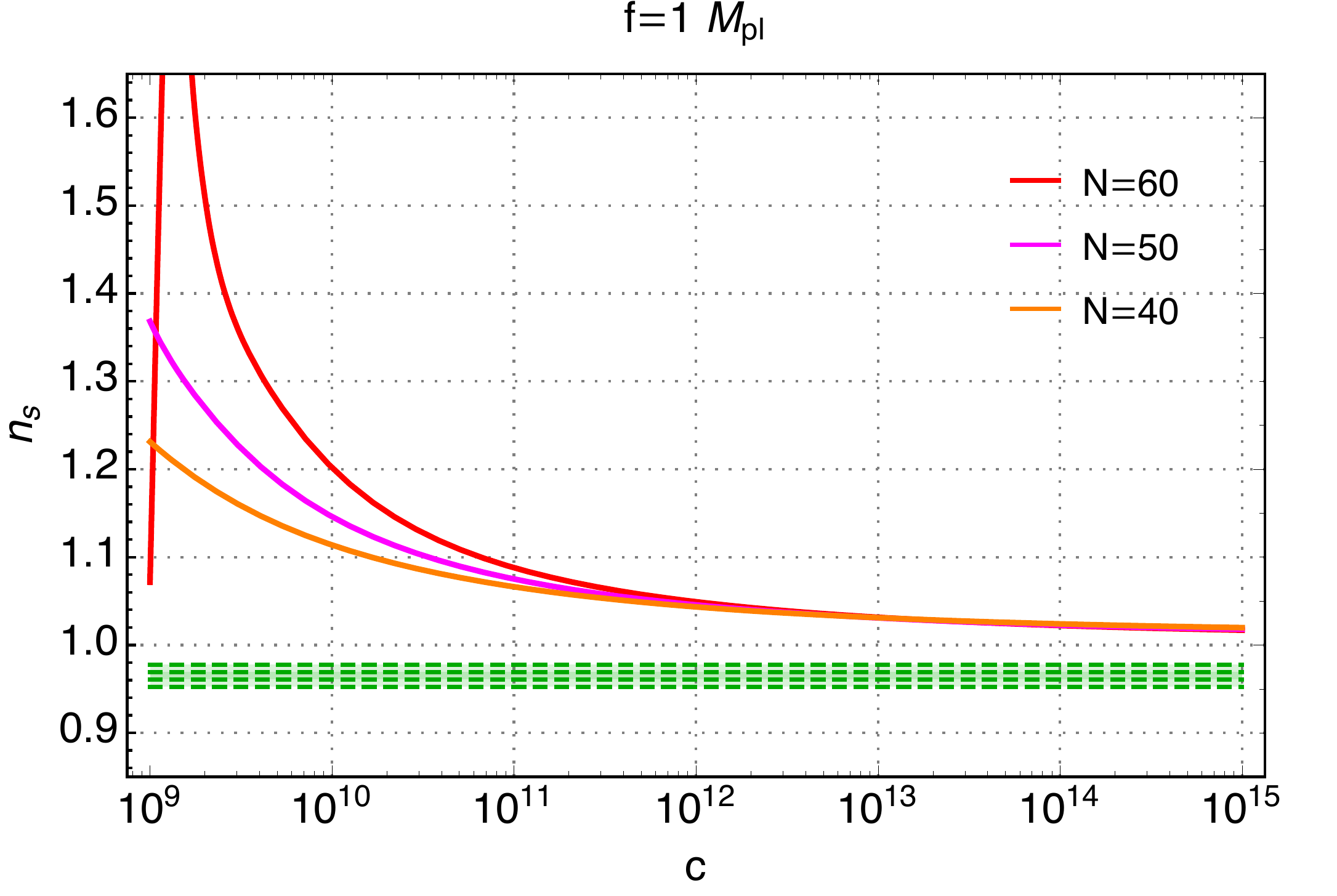}
\includegraphics[width=0.49\textwidth]{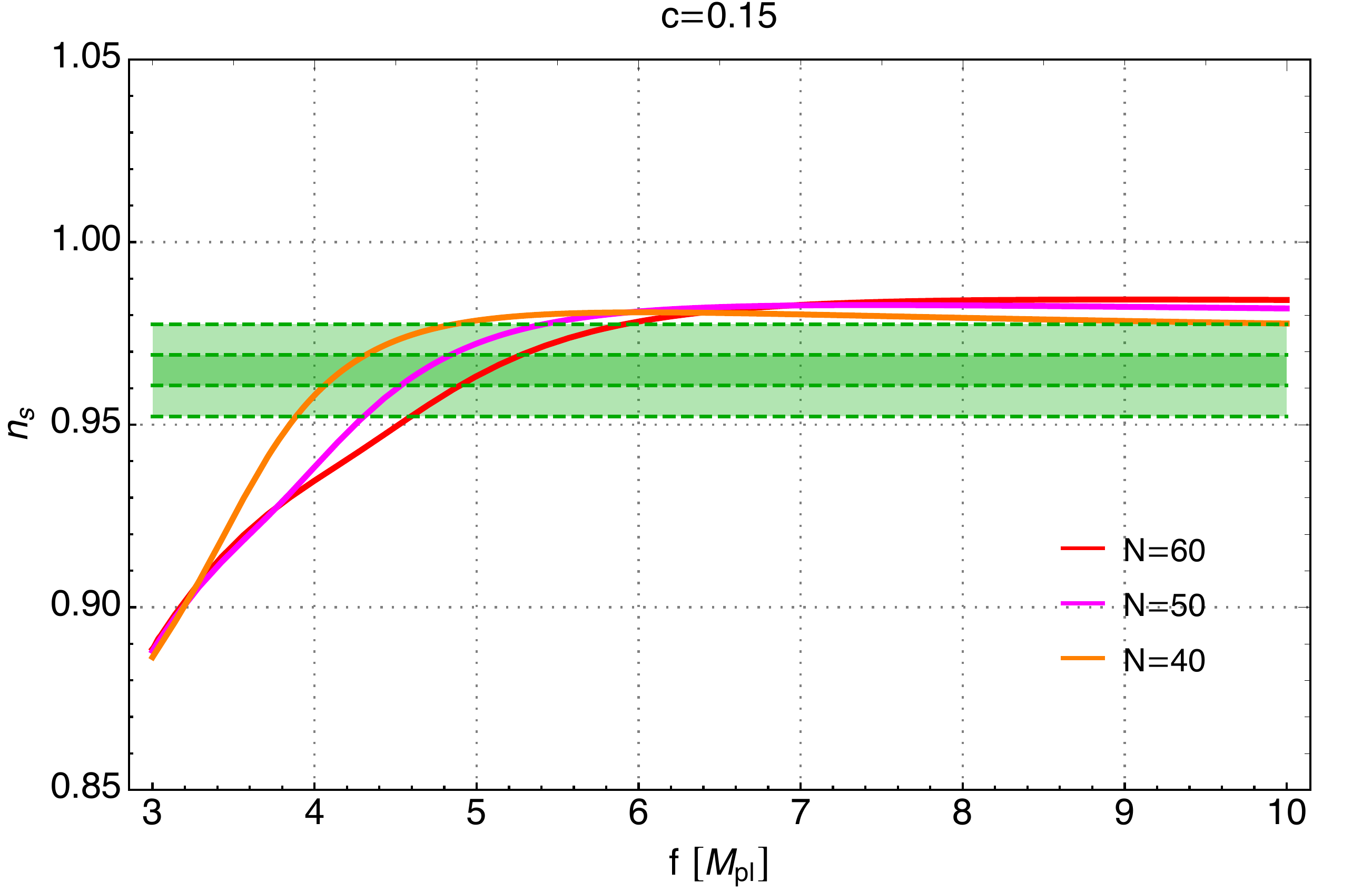}
\includegraphics[width=0.49\textwidth]{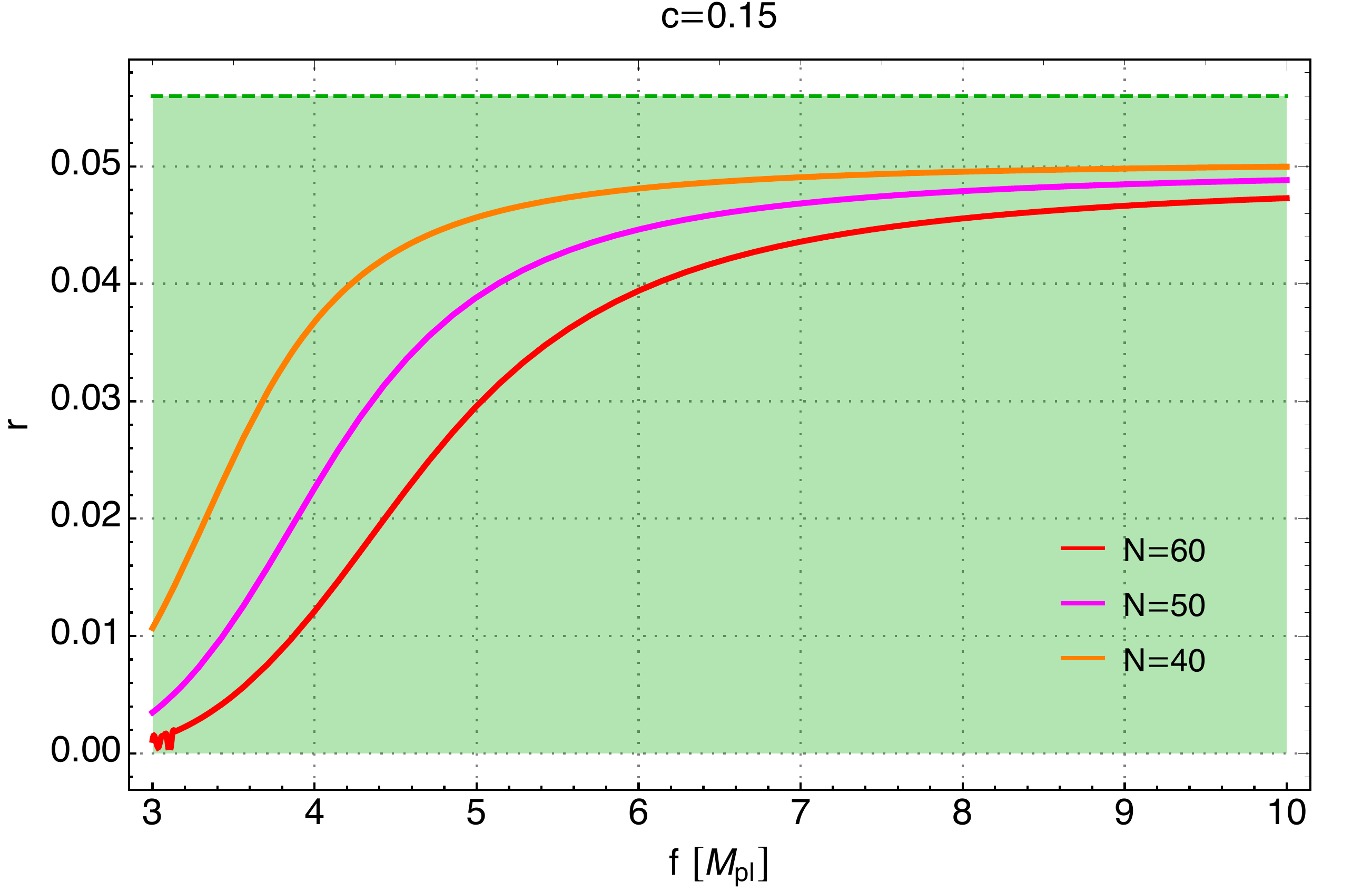}
\caption{Numerical results in case of thermalized inflaton towards a smaller $f$. First row: we fix $f=1M_{\rm pl}$. The left plot shows the number of e-folds with variations in $c$, where the horizontal dashed line marks $N=60$. The right plot shows the predicted spectral index as a function of $c$, where the green bands show the Planck-allowed region. Second row:  we fix $c=0.15$. The left plot shows the spectral index as a function of $f$, and the right plot shows the tensor-to-scalar ratio as a function of $f$. The green bands show the Planck-allowed region. }
\label{fig:smallf}
\end{figure}

We show our numerical result in $f=1M_{\rm pl}$ case in the first row of Fig.~\ref{fig:smallf}. Enough e-folds ($N \gtrsim 60$) require that $c\gtrsim 10^9$, which results in a $n_s \gtrsim 1$ -- outside the Planck allowed range. Actually, it is a generic feature that for large c, $n_s \gtrsim 1$. We have checked numerically for $f=0.1 M_{\rm pl}$ and $f=0.01 M_{\rm pl}$ cases, enough e-folds allow only large c, which also lead to $n_s \gtrsim 1$.

Is it possible to have a proper $n_s$ with other small values of $f$? To answer this, it is useful to show the spectral index as a function of $f$. To do this, we fix $c=0.15$, which is inspired from the $f=5M_{\rm pl}$ result and only serves for illustrative purpose. Our results are shown in the second row of Fig~\ref{fig:smallf}. From this figure, we see that the spectral index grows with $f$. The lower limit $f\gtrsim 3 M_{\rm pl}$ is set by requiring enough e-folds ($N \gtrsim 60$) and is only valid for $c=0.15$. We expect this lower limit may vary given other allowed $c$ value, which does not interest us here. Nevertheless, the exclusion of $f=1M_{\rm pl}$ case is generic in $c$, and the feature that $n_s$ grows with $f$ when $c$ is fixed should also be generic. Thus we conclude that cases with $f \leqslant 1M_{\rm pl}$ are excluded by the Planck results in our model.

\begin{figure}
\centering
\includegraphics[width=.7\textwidth]{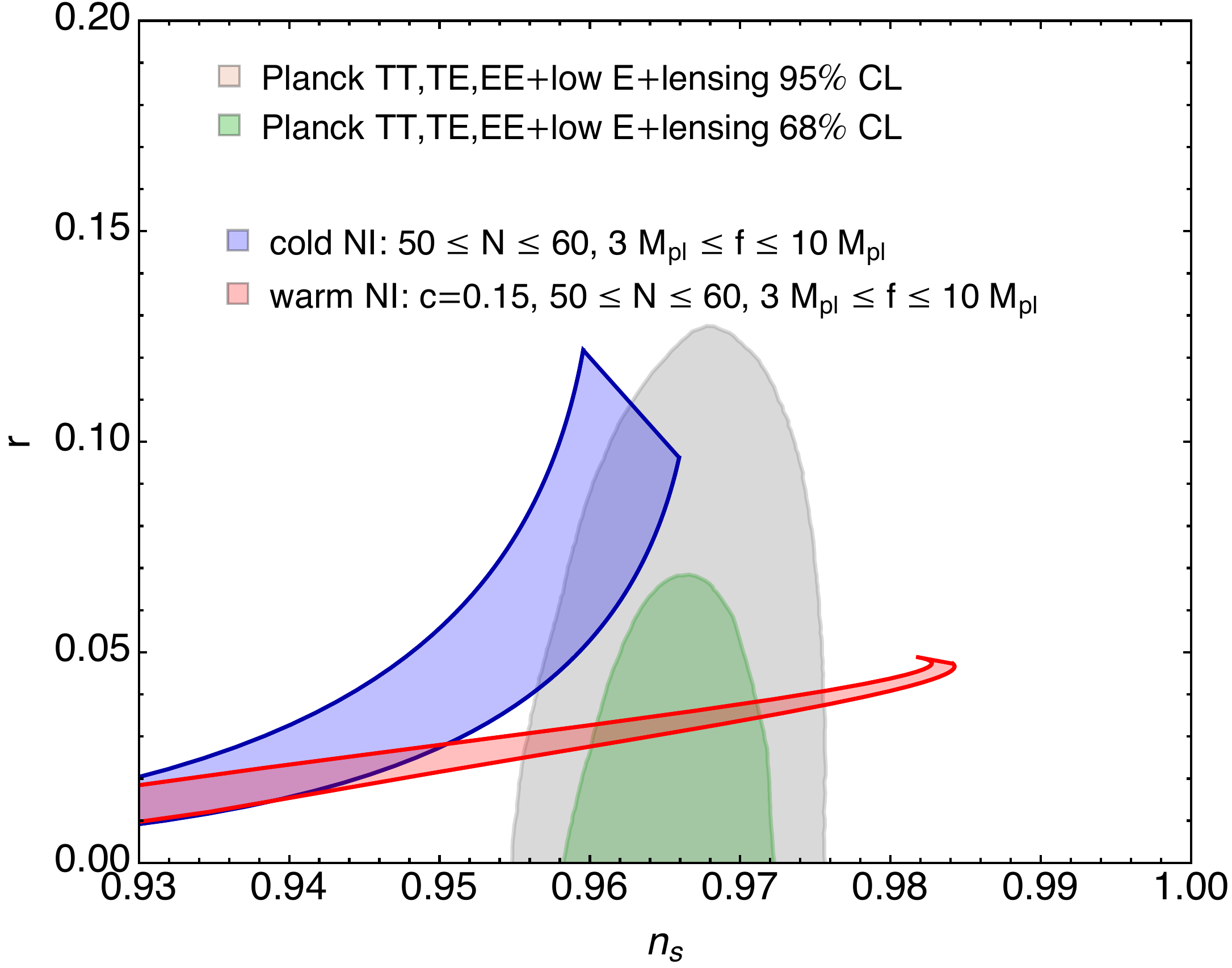}
\caption{ Comparison of the warm natural inflation ($c=0.15$) with cold natural inflation. We choose the same $f$ and $N$ range, and plot with the Planck 2018 constraints~\cite{Akrami:2018odb}. Here inflaton is thermalized.
}
\label{fig:wpc15}
\end{figure}

It is as well interesting to make a direct comparison with the cold natural inflation prediction, and we show it in Fig.~\ref{fig:wpc15}. We see the parameter space shrinks slightly due to the suppressed $r$, and $n_s$ is slightly blue-tilted. The combination of the two effects brings the cold inflation into the Planck-allowed region when warm dissipative effects are included.

The result here is different from that in Refs.~\cite{Mohanty:2008ab,Visinelli:2011jy,Mishra:2011vh} for mainly two reasons. First, in these works, the dissipative coefficient is independent of temperature, which results in $\beta_w = 0$. As $|\beta_w|$ grows with decreasing $f$ very quickly, it closes the slow-roll region in the small $c$ range very soon. This happens before we calculate e-folds. The value of $\beta_w$ rules out a smaller $f$ in the small $c$ end. For a large $c$ and small $f$, there exist regions for slow roll, but the resulting spectral tilt is in general larger than $1$ and is not viable. Second, Refs.~\cite{Mohanty:2008ab,Visinelli:2011jy,Mishra:2011vh} work in the strong dissipative regime and thus only thermal fluctuation is considered. In our work, with both thermal and quantum fluctuations, we see that only a weak dissipative effect is enough to bring natural inflation into agreement with observation.

\section{Conclusions} \label{sec:conclusion}

We consider natural inflation in a warm inflation framework with a dissipative coefficient having a temperature dependence as $\Gamma \propto  T^3$. Natural inflation can be compatible with the Planck 2018 results when such dissipative effects are included (see Figs.~\ref{fig:f5_3}, ~\ref{fig:nonthermal_2} and~\ref{fig:wpc15}). With no a priori assumptions on the magnitude of the dissipative effect, we find that the Planck results prefer a weak to an intermediate dissipative regime for our benchmark scale $f=5 M_{\rm pl}$ depending on whether inflaton is thermalized or not. On the other hand, $f=5 M_{\rm pl}$ lies outside the $2\sigma$ region in cold case. Moreover, inflation starts with quantum fluctuation domination and evolves with a growing thermal fluctuation that dominates over quantum fluctuation before the end of the inflation. The observed spectral tilt puts stringent constraints on the model's parameter space. For $c=0.15$ in case of thermalized inflaton, the decay constant is limited to $[4.6,6] M_{\rm pl} $ for at least $60$ e-folds. It is the first time natural inflation with a thermal dissipative effect is viable in the weak regime to our knowledge.

As is concerned when trying to embed natural inflation into a more fundamental theory, we discuss the possibility of having a decay constant that is smaller than the reduced Planck mass. We find that in such cases requiring enough e-folds leads to a large $c$, which results in a blue-tilted spectral index that has been excluded. Our result is different from that of the existing studies~\cite{Mohanty:2008ab,Visinelli:2011jy,Mishra:2011vh}, which work in the strong dissipative regime and take dissipative coefficient independent of temperature and allow a sub-Planckian $f$.

Such temperature dependence of the dissipative coefficient may stem from the inflaton being an axion-like particle couples to gauge fields, which has been justified in the strong dissipative regime and remains to be clarified in the weak regime~\cite{Berghaus:2019whh}. We find tension with the thermal field theory requirement for such origin of the dissipative coefficient. A non-thermalized inflaton relaxes the constraint on $\alpha$ and alleviates the tension but not eliminates it. Further studies on the thermal field theory estimation of this origin are desired to clarify this issue. Although we write the dissipative coefficient in this form (Eq.(\ref{eq:gamma})), our main results, as mentioned above, are valid as long as $\Gamma \propto  T^3$, and do not depend on the validity of the dissipative coefficient's origin in the weak regime.

Besides the predictions on the spectral index and the tensor-to-scalar ratio, when the dissipative term originates from the inflaton's coupling to gauge fields, a Q-independent prediction on the non-gaussianity is given as $f_{\rm NL}^{\rm warm} \approx 5$~\cite{Berghaus:2019whh}. This value can be further decomposed with the most constrained bispectral shape being $f_{\rm NL}^{\rm local} \approx 1.5$, which is within the Planck 2018 allowed range: $f_{\rm NL}^{\rm local} = -0.9 \pm 5.1$~\cite{Akrami:2019izv}. When the local non-gaussianitity is discovered, it is possible to conclusively test the model by its unique bispectrum shape~\cite{Bastero-Gil:2014raa}.

Besides our approach here, natural inflation can be brought into better agreement with the observations in other frameworks. For example, in a UV complete quadratic gravity framework~\cite{Salvio:2019wcp}, or with a nonminimal coupling to gravity~\cite{Reyimuaji:2020goi}. Being an attractive model itself, natural inflation deserves further studies. Distinctive features of different realizations can be confronted with future observations, which hopefully renders a unique answer regarding natural inflation.

\section*{Acknowledgement} 
X. Z. acknowledges helpful correspondence with Prof. Ramos. Y. R. is grateful to the support from the postdoctoral research fellowship of China, and X. Z. is supported partically by China Postdoctoral Science Foundation under Grant No. 2019M650001.

\end{document}